%% file: paper.tex
\documentclass[12pt]{article}
\usepackage{graphicx} 
\usepackage{amsmath}
\usepackage{dcolumn}
\usepackage{epsfig}
\usepackage{exscale}
\usepackage{color}
\input rotate
\input macro
\newcommand{\be}{\begin{equation}}
\newcommand{\ee}{\end{equation}}
\newcommand {\bm}[1]{\mbox{\boldmath$#1$}}

\newdimen\unit
\def\Xint#1{\mathchoice
{\XXint\displaystyle\textstyle{#1}}%
{\XXint\textstyle\scriptstyle{#1}}%
{\XXint\scriptstyle\scriptscriptstyle{#1}}%
{\XXint\scriptscriptstyle\scriptscriptstyle{#1}}%
\!\int}
\def\XXint#1#2#3{{\setbox0=\hbox{$#1{#2#3}{\int}$}
\vcenter{\hbox{$#2#3$}}\kern-.5\wd0}}

\def\dashint{\Xint-}
\def\position#1 #2 #3 {\rlap{\kern#2\unit
    \raise#3\unit\hbox{#1}}}
\definecolor{brown}{rgb}{0.59,0.29,0.0}
\begin{document}
\title{ Linear MHD stability studies with the 
~~~~~~~~~~~~~~~~~~~~~~~ STARWALL code}
\author{P.~Merkel, and  E.~Strumberger  \\
Max Planck Institute for Plasma Physics, Boltzmannstr. 2.  \\
\sl   85748 Garching, Germany}
\maketitle
\baselineskip = 18.0truept plus 0.5pt minus 0.5pt
\medskip             
\begin{abstract}
The STARWALL/CAS3D/OPTIM code package is a powerful
tool to study the linear MHD stability of 3D, ideal equilibria in the
presence of multiply-connected ideal and/or resistive conducting 
structures, and their feedback stabilization by external  currents.
Robust  
feedback stabilization of resistive wall modes
can be modelled with the OPTIM code.
Resistive MHD studies are possible combining STARWALL with the
linear, resistive 2D CASTOR code as well as nonlinear MHD simulations 
combining STARWALL with the  JOREK code.

In the present paper, a detailed description of the STARWALL code is given  
and some of its applications are presented to demonstrate the methods
used.
Conducting structures are treated in the thin wall 
approximation and  depending on their complexity  they
are discretized by a spectral method or by triangular finite elements.
As an example, a  configuration is considered consisting of an ideal plasma 
surrounded by a vacuum domain containing a resistive wall and bounded 
by an external wall.
Ideal linear MHD modes and resistive wall modes in the presence of multiply-connected walls are studied.
In order to treat the vertical mode self-consistently the STARWALL code 
has been completed by adding the so-called L\"ust-Martensen term 
generated by a constant normal displacement of the plasma.

The appendix contains the computation of the 2D Fourier transform of
singular inductance integrals, and the derivation of an asymptotic 
expansion for large Fourier harmonics. 
\end{abstract}
\par
\medskip
\par
\bigskip
\bigskip
\section{Introduction}
\vskip0.42truecm
\input introduction_m.tex
\section{The vacuum contribution}
\label{sec:solution}
\vskip.42truecm
\input vacuum.tex
\section{Variational method}
\input variation.tex
\vskip.42truecm
\section{The finite element method}
\label{sec:finite}
\input finitelem.tex
\section{Application}
\input kapitel_cor.tex
\label{sec:appli}
\input outlook_m.tex
\par
\bigskip
\bigskip
\centerline{\large \bf Acknowledgments}
\par
\medskip
We woul like to thank Sibylle G\"unter, Karl Lackner, and Matthias H\"olzl
for useful discussions and comments.
\par
\bigskip
\bigskip 
\addcontentsline{toc} {section} {REFERENCES}
\bibliography{reference}
\par
\newpage
\section{Appendix}
\begin{appendix}
\numberwithin{equation}{section}
\input lagrangian.tex
\section{Inductance matrices}
\label{sec:spectral}
\input inductance.tex

\section{The subtraction method }
\input subtract.tex
\section{The regularisation integrals }
\input regul_integrals.tex
\input asymptotic.tex
\end{appendix}
\end{document}

%% file: macro.tex
\newcounter{itemnum} \newlength{\addnum}

%% file: introduction_m.tex
\par
\medskip

{The STARWALL code has originally been developed for the numerical
treatment of resistive wall modes (RWMs). It has been applied to many different physics
problems via a coupling to linear and non-linear MHD codes already (see also the outlook
in Section 6). The present paper concentrates on STARWALL itself, in particular on the
mathematical and numerical methods used.}

{After briefly introducing resistive wall modes, this
introductory part explains the main features of the STARWALL code and its interaction with other
codes. Feedback stabilization studies of RWMs are described,
and the recently added L\"ust-Martensen term which is important for axisymmetric
instabilities is introduced. Finally the outline of the rest of the paper
is given.}

External kink unstable tokamak equilibria {which are fully} stabilized
by an ideal conducting wall sufficiently close to the plasma remain unstable {in
the presence of a a realistic wall} because of its finite resistivity. {These instabilities,}
Resistive Wall Modes (RWMs), grow on the time scale of the magnetic field diffusion through
the resistive wall.
{With the growth rate of the RWM} being typically orders of magnitude smaller
than {that of} the kink mode in the {absence of a wall,
stabilizing the modes with} an active feedback system becomes feasible.
The topical review on stabilization of the external kink and the resistive
wall mode by Chu and Okabayashi~\cite{Chu2010} {gives a comprehensive overview
of the existing literature on this topic.}

Beside STARWALL -- presented here -- several other codes have been developed and
used to study RWMs: VALEN \cite{Bialek2001,Bialek2007}, DCON coupled to VACUUM 
\cite{Chu2003}, MARS-F \cite{Liu2004}, CarMa \cite{Portone2008}.
In the VALEN code the plasma state is approximated by a single unstable 
eigenfunction, whereas in MARS-F and CarMa the stability of 2D equilibria
in the presence of 3D structures is treated. STARWALL is the only
code which can be applied to 3D equilibria with
general 3D wall configurations. {Both, VALEN and STARWALL
use a thin wall approximation.}

The STARWALL code is part of a comprehensive code package: 3D equilibrium NEMEC
code \cite{Hirshman1983,Hirshman1986,Hirshman1_1986}, coordinate transformation
COTRANS code, 3D ideal MHD stability CAS3D code \cite{Nuehrenberg1996}, vacuum
and RWM stability STARWALL code, and the feedback optimization OPTIM code \cite{Sempf2009}.

{An adapted version of the STARWALL code has also been coupled
\cite{Hoelzl2012} to the non-linear MHD code JOREK~\cite{Huysmans2007} to include resistive
wall effects in the simulations and has already been applied to a variety of different physics
problems. The CASTOR3D code is currently under development~\cite{Strumberger2014a},
for linear stability studies of 3D equilibria including 3D resistive wall effects described in the
STARWALL formalism. Some additional information on JOREK-STARWALL and CASTOR3D is provided in Section 6.}

Computations of external modes with ideal conducting wall configurations
{are possible with the CAS3D code} including the perturbed
kinetic energy {where the vacuum energy term is provided by} the STARWALL
code. In case of slowly growing resistive wall modes, the {plasma inertia}
can be neglected such that a problem of first order in time is obtained~\cite{Chu2003}.
{These modes can be studied with the STARWALL code. Also, the
feedback stabilization of RWMs can be investigated as described in the following.} 
The feedback procedure can be divided in two parts, the open-loop and the closed-loop
problem. In the open-loop part, a complete set of eigenfunctions of the plasma-resistive-wall 
system is determined. The feedback coils can be  included in the resistive wall configuration
 passive resistive elements without external voltage applied.

{The closed-loop part consists of feedback logics which calculate the optimal
voltages to be applied at the feedback coils based on signals which are produced by sensor
coils at appropriate positions.
The feedback code OPTIM~\cite{Sempf2009} implemented for this purpose achieves
robust control by considering} all unstable, but also a set of stable modes to assure
that modes are not driven unstable by the active feedback.

RWM kink modes and their feedback stabilization have been studied for
ITER and ASDEX Upgrade tokamak-type configurations  with realistic wall
structures and for 3D quasi-axisymmetric equilibria
 \cite{Merkel2004,Merkel2006,Guenter2008,Strumberger2008,Kallenbach2011a}.

To study external axisymmetric modes (toroidal harmonic $n=0$) the STARWALL 
code has been
completed  by implementing 
the so-called  L\"ust-Martensen term \cite{Martensen1960}.
The vacuum contribution of  an external mode is determined by the normal displacement $\xi_{n}$ at the plasma boundary.
For $\xi_n \ne  const$  the vacuum contribution is the solution of a Neumann-type
problem with prescribed normal component ${\bf B}_n$.
The   L\"ust-Martensen term is the plasma perturbation
for $\xi_n= const$ where the perturbed ${\bf B}_{vac}$ is tangential at the plasma boundary.
 In that case a net-toroidal and net-poloidal magnetic flux is induced in the vacuum region  
bounded by the plasma boundary  and an external ideal conducting  
wall   or at least by ideal  toroidal field coils.
That is, for  vertical mode studies the identical physical configuration should be used by which the free-boundary equilibrium was generated.
Results of Vertical Displacement Events (VDEs) taking into account
the L\"ust-Martensen term and their coupling to higher toroidal harmonics via
a three-dimensional resistive wall are presented for an AUG-type configuration in Section~5.

The {rest of this} paper is organized as follows. In Section 2,
the {physical} problem to be solved {for vacuum, resistive
wall, and ideal plasma} is defined. In Section 3 {the variational procedure to solve this
problem using a spectral discretization is described, and in Section 4 the method used
for triangular finite elements is introduced. Section 5 contains an example for applying
STARWALL to realistic geometries: Linear stability is investigated for
an ASDEX Upgrade-type configuration including a 3D resistive wall with holes.}
{In Section 6, a brief outlook to future applications of the STARWALL
code for linear and non-linear MHD problems is given.}

{In Appendix A the variational method is explained comprehensively,} and Appendix B
contains the  definition of the inductance matrices. In Appendix C the subtraction
method for the treatment of the singular matrix elements is described.
Appendix D contains the computation of the 2D Fourier transform of
singular inductance terms already published in \cite{Merkel1986} and
\cite{Merkel1987}. The treatment of unstable recurrence relations
by a method running the recursion in the backward direction has been added.
Furthermore, the derivation of an asymptotic expansion for large values of the 
$2D$ Fourier harmonics $m,n$  for the singular Fourier integrals is given. 

%% file: vacuum.tex
\par
\medskip
The STARWALL procedure will be explained considering an ideal plasma equilibrium in the presence of a multiply-connected resistive wall and an external ideal conducting (superconducting) wall
(see Fig.~1 and Fig.~2). For the resistive wall the  thin-wall approximation is used \cite{Freidberg1989}.
 Assuming
the RWM to be sufficiently slow, the kinetic energy of the plasma perturbation can be neglected.
In analogy to the ideal energy principle \cite{Bernstein1958} a variational technique can be applied.
The Lagrangian
\begin{equation} \label{200}
   {\cal L} = W_p(\bm {\xi},\bm {\xi}) +
{1 \over 2 } \int_{S_1}\! df\
({\bf n} \cdot {\bm {\xi}})  ({\bf B} \cdot {\bf B}_0) 
\end{equation}
consists of the 
 potential energy of the plasma 
perturbation and  the contribution of the vacuum domain  given by a surface integral
at the 
  plasma-vacuum interface ($S_1=$ plasma-vacuum interface, 
  ${\bm \xi}$ = displacement vector,
 ${\bf B}_0$= equilibrium magnetic field,
 ${\bf B}$= perturbed vacuum magnetic field,
 ${\bf n}$ =
  exterior normal on $S_1$).
 The perturbed magnetic field ${\bf B}$ is uniquely determined by the normal component
${\bf n} \cdot {\bm \xi} $ of the displacement ${\bm \xi}$ at $S_1$.

The potential energy of the plasma perturbation \cite{Nuehrenberg1987} - provided 
by the CAS3D code -
is given by
\begin{eqnarray}    \label{201}
W_{p}(\bm {\xi},{\bm \xi})  & = &\frac{1}{2} \int_{S_1} d^3r [ {\bf C}^2 + \Gamma p (\nabla \cdot \bm {\xi})
^2
   -{\cal A} (\bm {\xi} \cdot {\bf n})^2],  \\
  {\bf C} & = & \nabla \times (\bm {\xi} \times {\bf B}_0) +
   (\bm {\xi} \cdot {\bf n}) {\bf j} \times {\bf n},~~~~~~ 
{\cal A}  =  2 ({\bf j} \times {\bf n}) \cdot ({\bf B}_0 \cdot \nabla) {\bf n},  \nonumber  \\
{\bf B}_0 & = & {\nabla s} \times {\nabla ( F'_P~v-F'_T~u)},  \nonumber
\end{eqnarray}
where  $s,u,v$ are flux coordinates: 
   $s:0\le s \le 1$  is the normalized toroidal flux,
and  $(u,v): 0\le u,v \le 1$ are poloidal and toroidal magnetic
coordinates on the surfaces.
$F'_P,F'_T$ are the derivatives of the poloidal and toroidal flux with respect to $s$.
The Fourier expansion of the displacement vector reads
\begin{equation} \label{202}
{\bm {\xi}}(s,u,v) = \sum_{m=0,n=0 \atop m>0,n=-nf}^{mf,nf}
  \bm{\xi}_s(s)_{mn} \sin 2 \pi (mu +n  v)+
 \bm{\xi}_c(s)_{mn} \cos 2 \pi (mu +n  v).
\end{equation}
With respect to the  flux coordinate $s$, the Fourier harmonics 
${\bm{\xi}}_s(s)_{mn} ,{\bm{\xi}}_c(s)_{mn} $ are discretized 
using  a finite element method.

The perturbed magnetic field ${\bf B}$  has
 to satisfy Maxwell's equations
\begin{equation}  \label{203}
         { \bf B} = \nabla \times {\bf A} ,~~
{\bf \nabla} \times ({\bf \nabla} \times {\bf A} ) = 0,~~~~
{\bf \nabla} \cdot {\bf A}   = 0 
\end{equation}
with
   boundary conditions for the vector potential ${\bf A}$ at the plasma-vacuum interface, the resistive wall and the external   superconducting wall.

On  the resistive wall the boundary condition  follows from Faraday's and
Ohm's law:
$
 {\bf E} + {\partial {\bf A } \over \partial t} = 0 ,~
\sigma {\bf E }= {\bf J}
$
.
Assuming a time dependence $e^{\gamma t}$, one gets in the thin wall
approximation the boundary conditions for the perturbed vector potential 
\begin{equation}   \label{204}
{\bf n}\times {\bf A}  = \left\{ \begin{array} {r@{\quad:\quad}l}
-({\bf n} \cdot {\bm {\xi}}) {\bf B}_{0} &
   \mbox{ on~~ $S_{1}$  (plasma-vacuum~ interface)} \\
   - \frac{1}{\sigma d \gamma} {\bf n} \times {\bf j}_2 & \mbox{     on~~ $S_{2}
$  (resistive~ wall )}  \\
  0 & \mbox{ on~~ $S_{3}$  (external conducting wall)}
\end{array} \right.
\end{equation}
where ${\bf j}_2$ is the  current in  the resistive wall,~
${\bf n}$ is the exterior normal unit vector on the surfaces, 
 $\sigma d$ is the surface resistance of the wall 
 ($\sigma=$ conductivity, $d=$ wall thickness)
and ${\bf B}â_0$ is the equilibrium magnetic field.

With the contravariant normal component $\nabla s \cdot {\bm \xi}$
used in the CAS3D code the boundary condition at the plasma-vacuum interface
reads 
\begin{equation} \label{205}
{\nabla s} \times {\bf A}  = 
-({\nabla s} \cdot {\bm {\xi}}) {\bf B}_{0},~~~~~ \nabla s = \frac{\mid {\bf r}_{v}   \times {\bf r}_u \mid}{
                 ({\bf r}_{v} \times {\bf r}_{u}) \cdot {\bf r}_{s}} {\bf n}, ~~~
     {\bf B}_0 = \frac{F'_T~ {\bf r}_{v}+  F'_P ~{\bf r}_{u}}{
                      ( {\bf r}_{v} \times {\bf r}_{u}) \cdot {\bf r}_{s}}
\end{equation}
 with   the equilibrium field ${\bf B}_{0}$  in magnetic coordinates 
 and ${\bf r}_u  := \frac{\partial {\bf r}}{\partial u},
  {\bf r}_v  := \frac{\partial {\bf r}}{\partial v},
  {\bf r}_s  := \frac{\partial {\bf r}}{\partial s}$.

Multiplying  (\ref{204})  with $ {\bf n} \times  {\bf r}_u$ and 
 $ {\bf n} \times {\bf r_v}$  one gets 
\begin{eqnarray} \label{206}
 {\bf r}_{1,u}  \cdot {\bf A} & =&  F'_T ~~\nabla s \cdot {\bm \xi}, ~~~~~~~~  
 {\bf r}_{1,v}  \cdot {\bf A}  =  -F'_P~~ \nabla s \cdot {\bm \xi} 
 ~~~~~ : ~~ \mbox{ on $S_1$} \nonumber \\ 
  {\bf r}_{2,u}  \cdot {\bf A} &=& 
   - \frac{1}{\sigma d \gamma} {\bf r}_{2,u} \cdot {\bf j}_{2}, ~~~~
  {\bf r}_{2,v}  \cdot {\bf A} = 
   - \frac{1}{\sigma d \gamma} {\bf r}_{2,v} \cdot {\bf j}_{2} 
 ~~~~ : ~~ \mbox{ on $S_2$ }\\ \nonumber
 {\bf r}_{3,u}  \cdot {\bf A} & =&  0,  ~~~~~~~~~~~~~~~~~~~~
 {\bf r}_{3,v}  \cdot {\bf A}  =  0~~~~~~~~~~~~~~~~~
 ~~ : ~~ \mbox{ on $S_3$} \nonumber 
\end{eqnarray}
        
The boundary conditions for the  perturbed magnetic field ${\bf B}$ are  obtained
by taking the divergence:
[$ \nabla \cdot (\nabla s \times {\bf A}) = {\bf A} \cdot( \nabla \times \nabla s) - \nabla s \cdot  {\bf B} $].
At the plasma-vacuum interface $S_1$  one gets
\begin{equation}  \label{207}
 \nabla s \cdot {\bf B}  = 
{\bf B}_{0} \cdot \nabla  (\nabla s   \cdot {\bm {\xi}})  
\end{equation}
 or
\begin{equation} \label{208}
{\bf N} \cdot {\bf B}  = \left\{ \begin{array} {r@{\quad:\quad}l}
      F'_T~ ({\nabla s} \cdot {\bm \xi})_v
      +F'_P~({\nabla s} \cdot {\bm \xi})_u &
   \mbox{ on~~ $S_{1}$  (plasma-vacuum~ interface)} \\
    \frac{1}{\sigma d \gamma} ({\bf r}_{2,v} \cdot {\bf j}_{2,u}
   -{\bf r}_{2,u} \cdot {\bf j}_{2,v})
 & \mbox{     on~~ $S_{2}
$  (resistive~ wall )}  \\
  0 & \mbox{ on~~ $S_{3}$  (external conducting wall)}
\end{array} \right.
\end{equation}
with  ${\bf N} = {\bf r}_v \times {\bf r}_u$  to be taken at the surfaces.

Prescribing the normal component ${\nabla s} \cdot {\bf B}$ of ${\bf B}$ one has to solve a Neumann-type
problem except for  
$\nabla s   \cdot {\bm {\xi}}=constant $  where the normal  component 
of  ${\bf B}$ vanishes. To get the boundary condition for this case
one has to go back to the boundary condition for the vector potential ${\bf A}$.

The solution for the vector potential ${\bf A}$ can be generated by surface currents on the
plasma-vacuum interface $S_1$, the resistive wall $S_2$ and the external ideal
conducting 
wall $S_3$  
\begin{equation} \label{209}
{\bf A}({\bf r})= {1 \over 4\pi}\sum_{i=1}^3 
 \int \limits_{S_i} df_i \
{ {\bf j}_i  \over \mid {\bf r}-{\bf r}_i \mid}
\end{equation}
    
 with divergence-free surface currents  derived from  current potentials
$\Phi_i$
\begin{equation} \label{210}
{\bf j}_i=  {\bf n }_i \times \nabla \Phi_i, ~~ {\bf n}_i 
= \frac{{\bf r}_{i,v} \times {\bf r}_{i,u}}{ \mid {\bf r}_{i,v} \times {\bf r}_{i,u} \mid},~~ i=1,2,3
\end{equation} 
 where $ {\bf n }_i,~i=1,2,3$ are the exterior unit normal vectors.

For  toroidally and poloidally closed tori the current potential is given by
\begin{equation}  \label{211}
\Phi_i =  I^T_i~u+I^P_i~v + \phi_i(u,v),~~~~i=1,2,3
\end{equation}
where $I^T_i,I^P_i$ are  net-toroidal and net-poloidal currents on the torus.
They play a role  only if the diplacement ${\bm \xi}$ contains a component
 $\nabla s \cdot {\bm \xi} = $ constant  \cite{Martensen1960}.
 The  $\phi_i(u,v)$ are single-valued  potentials which
 can be expanded in Fourier space
for smooth tori 
\begin{equation} \label{212}
 \phi_i(u,v) = \sum_{m=0,n=1 \atop m>0,n=-n_c}^{m_c,n_c}
 \hat \phi^s_i(m,n)~ \sin 2 \pi (mu +n  v)+
\hat \phi^c_i(m,n)~ \cos 2 \pi (mu +n  v),~~  i=1,2,3 
\end{equation}

The surface currents have to be determined such that the boundary conditions for
${\bf A} $ on $S_i$, i=1,2,3 are satisfied. That will be carried out by means of
a variational procedure.

%% file: variation.tex
\par
\medskip
To treat  cases with multiply-connected resistive wall  configurations, a finite element 
method has been applied using a variational procedure \cite{Freidberg1989}.
One introduces the Lagrangian
\begin{eqnarray} \label{301}
{\cal L}_V\!\!\! & =&\!\! \!\frac{1}{8\pi} Ÿ\sum_{i=1}^3 \sum_{k=1}^3 \int \limits_{S_i} \! df_i \int \limits_{S_k} \!df_k
\frac{{\bf j}_i \cdot {\bf j}_k} {\mid {\bf r}_i-{\bf r}_k \mid}
+ \!\frac{1}{2 \gamma}  \int \limits_{S_2} \!df_2 \
\frac{{\bf j}_2 \cdot {\bf j}_2 }{\sigma d} 
\!+\int \limits_{S_1} \!df_1~({\bf n}_1\! \cdot \! {\bm {\xi}})~ {\bf n}_1\!
  \cdot \! ({\bf j}_1 \times {\bf B}_0) \nonumber \\ 
& & 
\end{eqnarray}

In appendix A it is shown that the first variation $\delta {\cal L}_V = 0$
gives the  boundary conditions (\ref{104a}),(\ref{105a}),(\ref{106a}).
Assuming a smooth plasma-vacuum interface being  identical with the
plasma boundary the current potential and the normal component of the displacement vector can be expanded in Fourier space
\begin{equation} \label{3001}
{\bm {\xi}} \cdot \nabla s = \sum_{m=0,n=0 \atop m>0,n=-n_f}^{m_f,n_f}
 \hat {\xi}_s(m,n) \sin 2 \pi (mu +n  v)+
 \hat{\xi}_c(m,n) \cos 2 \pi (mu +n v)   
\end{equation}
so that  one gets for
the last term of the Lagrangian ${\cal L}_V$ 
\begin{eqnarray} \label{3002}
 \int \limits_{S_1} \!df_1~({\bf n}_1\! \cdot \! {\bm {\xi}})~ {\bf n}_1\! \cdot \! ({\bf j}_1 \times {\bf B}_0)
 &=&
 - (F_T^\prime I^P_1 +F_P^\prime I^T_1)~\hat \xi_c(0,0) \\
 &+& \!\!\!\!  
 \sum_{m=0,n=1 \atop m>0,n=-n_f}^{m_f,n_f}\!\!\!\!\!\!\!\!
\pi( n F_T^\prime + m F_P^\prime) (\hat\xi_s(m,n) \hat \phi^c_1(m,n)
 -\hat \xi_c(m,n) \hat \phi_1^s(m,n))  \nonumber
\end{eqnarray}
 being  linear in $\Phi_1$ and determining the boundary condition at the plasma-vacuum interface.
 The variables of the Lagrangian ${\cal L}_V$ are the net-currents $I^T_i$,$I^P_i$ and the Fourier harmonics of the single-valued potentials $\hat\phi_i,i=1,2,3$. 
Replacing in ${\cal L}_V$ the currents using (\ref{210}) one gets 
\begin{equation} \label{302}
{\cal L}_V \!  = \!\frac{1}{2} Ÿ\sum_{i=1}^3 \sum_{k=1}^3 
 \hat\Phi_i^\top {\bf M}_{ik} \hat\Phi_k +\frac{1}{2 \gamma} ~ \hat\Phi_2^\top {\bf N}_{22} \hat \Phi_{2}
           + \hat\Phi_1^\top {\bf M}_{1 \hat \xi} \hat \xi           
\end{equation}
 with 
\begin{eqnarray} \label{303}
\hat \Phi_i^\top &=& [I_i^T,I_i^P,  \hat  \phi_i^s(0,1),...,\hat\phi_i^s(m_{f_i},n_{f_i}), 
          \hat \phi_i^c(0,1),...,\hat \phi_i^c(m_{f_i},n_{f_i})] \nonumber \\
\hat \xi^{ \top}  &=& [\hat \xi_s(0,1),...,\hat\xi_s(m_{f},n_{f}), 
          \hat   \xi_c(0,0),...,\hat \xi_c(m_{f},n_{f})] 
\end{eqnarray}
and  $\top \equiv$ transpose.

Varying  ${\cal L}_V$  with respect to $\hat \Phi_i,i=1,2,3$  one gets a set of linear equations for $\hat \Phi_i$ .
\begin{eqnarray} \label{304}
      {\bf M}_{pp}~{\hat\Phi}_p+ {\bf M}_{pw}~{\hat\Phi}_w +{\bf M}_{pt}~\hat\Phi_t  &= &-{\bf M}_{p \hat \xi}~\hat\xi \nonumber \\
      {\bf M}_{wp}~{\hat\Phi}_p+ {\bf M}_{ww}~{\hat\Phi}_w +{\bf M}_{wt}~\hat\Phi_t  &= &  0, \\
      {\bf M}_{tp}~{\hat\Phi}_p+ {\bf M}_{tw}~{\hat\Phi}_w +{\bf M}_{tt}~\hat\Phi_t  &= & -\frac{1}{\gamma} {\bf N}_{tt} ~\hat \Phi_t  \nonumber
\end{eqnarray}
(note: the index notation has been changed $p \equiv 1$ - plasma-vacuum interface, $t \equiv 2$ - resistive wall, $w \equiv 3$ - external conducting wall)

  The vacuum contribution in (\ref{200}) is given by
\begin{eqnarray} \label{305}
 W_{s}  &=& 
 \frac{1}{2 } \int_{S_1} du~dv ({\bm \xi} \cdot \nabla s)({F_P^\prime~ {\bf r}_{1,u}
+ F_T^\prime~ {\bf r}_{1,v}  }) \cdot {\bf B}
\end{eqnarray} 
with the magnetic field  generated by the currents on the surfaces $S_i,i=1,2,3$
\begin{equation} \label{306}
{\bf B}({\bf r})  = \frac{1}{4 \pi} \sum \limits_{i=1}^3 \left [ \int_{S_i}  du^\prime~dv^\prime
 \frac{(I_i^P~{\bf r}_{i,u^\prime}- I_i^T~{\bf r}_{i,v^\prime}) \times ({\bf r}-{\bf r}_i^\prime)}{|{\bf r}-{\bf r}_i^\prime|^3}
+  \nabla  \int_{S_i}( {\bf df}^\prime \cdot \nabla^\prime 
     \frac{1}{|{\bf r}-{\bf r}_i^\prime|})~ \phi_i^\prime  \right ]
\end{equation}
Inserting (\ref{306}) into (\ref{305}), $W_{s}$ splits into two terms
\begin{equation} \label{307}
 W_{s}  =   
  \frac{1}{2 } \sum \limits_{i=1}^3 (W_i^{(1)} +W_i^{(2)})
\end{equation}
 The contributions for $i=1$ are given by
\begin{eqnarray}
 W_1^{(1)}  &= &  \frac{1}{4 \pi}
 \int_{S_1} \!du~ dv 
 ({\bm \xi} \cdot \nabla s) \Big (
   \int_{S_1}\!  du^\prime dv^\prime    
 (F_P^\prime~ {\bf r}_{1,u}
+ F_T^\prime~ {\bf r}_{1,v}) \times
 (I_1^P~{\bf r}_{1,u^\prime}- I_1^T~{\bf r}_{1,v^\prime}) \cdot \frac{({\bf r}_1-{\bf r}_1^\prime)}{|{\bf r}_1-{\bf r}_1^\prime|^3} \nonumber \\ 
 & &~~~~~~~~~~~~~~~~~~~~~~~~~~~  + 2 \pi (F^\prime_P~I^T_1+F^\prime _T~I^P_1) \Big ) \nonumber \\
 W_1^{(2)}  &= & -\frac{1}{4 \pi}
 \int_{S_1} \!du~ dv 
\Big  [(F_P^\prime \frac{\partial}{\partial u}
  +F_T^\prime \frac{\partial }{\partial v})
 ({\bm \xi} \cdot \nabla s) \Big ] \Big ( 2 \pi~ \phi(u,v) \\ 
 & & ~~~~~~~~~~~~~~~~~~~~~~~~~~~~~~~~~~~~~~~~~+\dashint_{S_1}\!  du^\prime dv^\prime    
  ({\bf r}_{1,v^\prime} \times {\bf r}_{1,u^\prime}) \cdot \frac{({\bf r}_1-{\bf r}_1^\prime)}{|{\bf r}_1-{\bf r}_1^\prime|^3}~ \phi_1(u^\prime,v^\prime ) \Big )  \nonumber
\end{eqnarray}

  For $i=2,3$  the contributions to (\ref{307}) are given by
\begin{eqnarray}
 W_i^{(1)}  &= &  \frac{1}{4 \pi}
 \int_{S_1} \!du~ dv 
 ({\bm \xi} \cdot \nabla s)
   \int_{S_i}\!  du^\prime dv^\prime    
 (F_P^\prime~ {\bf r}_{1,u}
+ F_T^\prime~ {\bf r}_{1,v}) \times
 (I_i^P~{\bf r}_{i,u^\prime}- I_i^T~{\bf r}_{i,v^\prime}) \cdot \frac{({\bf r}_1-{\bf r}_i^\prime)}{|{\bf r}_1-{\bf r}_i^\prime|^3}  \nonumber\\
 W_i^{(2)}  &= & \frac{-1}{4 \pi}
 \int_{S_1} \!du~ dv 
\Big  [(F_P^\prime \frac{\partial}{\partial u}
  +F_T^\prime \frac{\partial }{\partial v})
 ({\bm \xi} \cdot \nabla s) \Big ]
   \int_{S_i}\!  du^\prime dv^\prime    
  ({\bf r}_{i,v^\prime} \times {\bf r}_{i,u^\prime}) \cdot \frac{({\bf r}_1-{\bf r}_i^\prime)}{|{\bf r}_1-{\bf r}_i^\prime|^3}~ \phi_i^\prime 
\end{eqnarray}
The $u,v$-integrations  have to be performed infinitesimal exterior of the 
plasma-vacuum interface $S_1$ and the singular integrals 
 $W^{(1)}_1$, $W^{(2)}_1$ are treated using a subtraction 
method(see appendix B).

With the definitions  (\ref{303})  the vacuum contribution $W_s$ reads
\begin{equation} \label{308}
 W_{s}  =  
  \frac{1}{2  } (\hat  \xi^{ \top }~ {\bf M}_{\hat \xi p}~ \hat \Phi_p  
  +\hat \xi^{ \top} {\bf M}_{i\hat \xi w}~ \hat \Phi_w  
  +\hat \xi^{ \top}~ {\bf M}_{\hat \xi t}~ \hat \Phi_t  )
\end{equation}
The  elements of the matrices ${\bf M}_{\hat \xi p},{\bf M}_{\hat \xi w}$
and ${\bf M}_{\hat \xi t}$ are given in appendix B. 

The variables of the system considered are the Fourier harmonics of the displacement vector
  $\bm {\xi}$, the current potentials
on the plasma-vacuum interface, the resistive wall and
the external wall. 
The set of equations (\ref{304}) determines the current potentials for  given
normal component 
$\hat \xi$ of the displacement vector  on the plasma-vacuum interface.

The system is closed by the equation derived from the Lagrangian (\ref{200}).
Varying the discretized functional (\ref{200}) one  obtains

\begin{equation}   \label{309}
 \left( \begin{array}{cc}
          {\bf M}_{XX} & {\bf M}_{X \hat \xi}   \\
          {\bf M}_{\hat\xi X} & {\bf M}_{\hat \xi \hat \xi}   \\
\end{array} \right) \cdot
\left( \begin{array}{c} X \\ \hat \xi
\end{array} \right)
-   \left( \begin{array}{c} 0 \\  {\bf M}_{\hat \xi p}~  \hat  \Phi_p+ {\bf M}_{\hat \xi w}~\hat \Phi_w + {\bf M}_{\hat \xi t}~ \hat \Phi_t 
\end{array} \right)  = 0.
\end{equation}

 The  variables are denoted as follows:
 $X=\{ {\bm{\xi}_s(s_i)}_{mn},\ldots,{\bm{\xi}_c(s_i)}_{mn},\ldots \}$ consists
 of all
components of
the displacement vector   except for  the Fourier harmonics of the
normal component  at the plasma boundary which are
denoted by $\hat \xi = \{\nabla s \cdot \bm{\xi}_s(1)_{mn},\ldots,\nabla s \cdot\bm{\xi}_c(1)_{mn},\ldots\}$. 

Solving the set of equations for $\hat \xi$   one gets

\begin{equation}  \label{310}
 \hat \xi   ={ \stackrel{cas3d}{{\bf M}_{\hat \xi \hat \xi}}}~( {\bf M}_{\hat \xi p}~ \hat\Phi_p +{\bf M}_{\hat \xi w
}~\hat \Phi_w + {\bf M}_{\hat \xi t}~\hat \Phi_t )
\end{equation}
with
\begin{equation}  \label{310a}
{ \stackrel{cas3d}{{\bf M}_{\hat \xi \hat \xi}}}~=({\bf M}_{\hat \xi \hat \xi}~
 -{\bf M}_{\hat \xi X}
     {\bf M}_{XX}^{-1}~ {\bf M}_{X \hat \xi})^{-1}   
\end{equation}

Eliminating $\hat \xi$ in (\ref{304}) one obtains
\begin{eqnarray} \label{311}
    \widetilde{\bf M}_{pp}~{\Phi}_p+ \widetilde{\bf M}_{pw}~{\Phi}_w+ \widetilde{\bf M}_{pt}~\Phi_t & = & 0, \nonumber \\
    {\bf M}_{wp}~\hat \Phi_p+ {\bf M}_{ww}~\hat \Phi_w+ {\bf M}_{wt}~\hat \Phi_t & = & 0, \\
      {\bf M}_{tp}~\hat\Phi_p+ {\bf M}_{tw}~\hat \Phi_w+ {\bf M}_{tt}~\hat \Phi_t & = &  
          - \frac{1}{\sigma d \gamma} {\bf N}_{tt}~\hat \Phi_t.    \nonumber 
\end{eqnarray}
 with
\begin{eqnarray} \label{312}
  \widetilde{\bf M}_{pp}&=&{({\bf M}_{pp}+ {\bf M}_{p \hat\xi}~{\stackrel{cas3d}{{\bf M}_{\hat \xi \hat \xi}}} {\bf M}_{\hat \xi p})}  \nonumber \\
  \widetilde{\bf M}_{pw}&=&{({\bf M}_{pw}+ {\bf M}_{p \xi}~{\stackrel{cas3d}{{\bf M}_{\hat \xi \hat \xi}}} {\bf M}_{\hat \xi w})} \\
  \widetilde{\bf M}_{pt}&=&{({\bf M}_{pt}+ {\bf M}_{p \hat\xi}~{\stackrel{cas3d}{{\bf M}_{\hat \xi \hat \xi}}} {\bf M}_{\hat \xi t})} \nonumber
\end{eqnarray}
Finally 
 one gets a set of linear equations defining an initial  
value problem of first order in time. 
Assuming a time dependence $e^{\gamma t}$, the normal modes of the system
are obtained by solving the generalized eigenvalue problem

\begin{eqnarray}  \label{313}
    \gamma      \left [ {\bf M}_{tt}-  
 \left( \begin{array}{cc}
          {\bf M}_{tp} & {\bf M}_{tw}   \\
\end{array} \right)  
 \left( \begin{array}{cc}
          \widetilde{\bf M}_{pp} & \widetilde{\bf M}_{pw}   \\
          {\bf M}_{wp} & {\bf M}_{ww}   \\
\end{array} \right)^{-1}  
 \left( \begin{array}{c}
          \widetilde{\bf M}_{pt}   \\
          {\bf M}_{wt}   \\
\end{array} \right) \right] 
 \hat \Phi_t  & = & - \frac{1}{\sigma d}  {\bf N}_{tt}~ \hat \Phi_t 
\end{eqnarray}

%% file: finitelem.tex
For the finite element procedure the conducting wall 
 is discretized into triangles.
The position vector of a triangle is given by
\begin{equation} \label{400}
 {\bf r}  = {\bf r}_1 + \alpha~ {\bf r}_{2,1} + \beta ~{\bf r}_{3,1}, ~~
        \alpha \ge 0,~ \beta \ge 0,~
        0 \le \alpha + \beta \le 1,~~
          {\bf r}_{i,k} = {\bf r}_i - {\bf r}_k,~~ i,k=1,2,3
\end{equation}
with the vertices numbered anti-clockwise $({\bf r}_1,{\bf r}_2,{\bf r}_3)$
and ${\bf r}_4 \equiv {\bf r}_1$. 

 Assuming the surface-current density  to be constant on the triangle 
 the current can be written as
\begin{equation} \label{401}
   {\bf j}_{\Delta} = \frac{ 
            \phi_1~{\bf r}_{2,3} +\phi_2~{\bf r}_{3,1}
  + \phi_3~{\bf r}_{1,2}}{ 
     \vert{\bf r}_{2,1} \times {\bf r}_{3,2} \vert}~~~
~~\mbox{or}~~~
{\bf j}_{\Delta} = \sum_{i=1}^{3} \phi_i~ {\bf e}_i,
    ~~{\bf e}_i = {1 \over 2} \varepsilon_{ikl} ~\frac {{\bf r}_{k,l} 
   } { \vert{\bf r}_{2,1} \times {\bf r}_{3,2} \vert}
\end{equation}
where the $\phi_i$ are the values of the current potential at the vertices of
the triangles.

The contribution of a pair of triangles to the functional ${\cal L}$  is 
given by
\begin{equation} \label{402}
 {\cal L}_{\Delta^{\prime} \Delta}=  
{\bf j}_{\Delta^\prime}\cdot {\bf j}_{\Delta}
  \frac{1}{8\pi}
\int_{\Delta^\prime} df' \int_{\Delta} df\
\frac{1} {\mid {\bf r}'-{\bf r}\mid} 
\end{equation}
\begin{equation*}
 {\cal L}_{ \Delta^{\prime} \Delta}=  
 \frac{1}{2} \sum_{i,k=1}^{3} \phi'_i L_{ik} {\phi_k},~~~~~~~~~~~
 L_{ik} =  {\bf e}^\prime_{i}\cdot {\bf e}_k ~\frac{1}{4 \pi} \
   \int_{\Delta^\prime} df' \int_{\Delta} df\
 \frac{1} {\mid {\bf r}'-{\bf r} \mid}
\end{equation*}
The  fourfold integral $L_{i,k}$  can be computed analytically.
 One gets
a sufficiently  good  approximation by performing two integrations analytically
and
the remaining two integrations numerically:

One obtains for the twofold integral
{
\begin{eqnarray} \label{403}
 \int_{\Delta^\prime}~df^\prime \frac{1}{|{\bf r} - {\bf r}^{\prime}|} & = &
 \sum_{i=1}^3 {|{\bf r}_{i+1,i}|}
  \Big ( a_i 
\ln \frac{l_{i}^++l_{i}^-+1}{l_{i}^++l_{i}^--1} \\
& -& h_i \Big
(\arctan \frac{a_i~ d_i^+}{a_i^2+h_i^2+ l_i^+ h_i} -
\arctan \frac{a_i~ d_i^-}{a_i^2+h_i^2+ l_i^- h_i} \Big) \Big) \nonumber
\end{eqnarray}
with
\begin{eqnarray*}
 l_i^{+}&=& \frac{|{\bf r}_{i+1}-{\bf r}|}{|{\bf r}_{i+1,i}|},~~
  l_i^{-}= \frac{|{\bf r}_{i}-{\bf r}|}{|{\bf r}_{i+1,i}|} ,~~
d_i^+  =  \frac{({\bf r}_{i+1} -{\bf r}) \cdot {\bf r}_{i+1,i}}{|{\bf r}_{i+1,i}
|^2},~~
d_i^- =\frac{({\bf r}_i -{\bf r}) \cdot {\bf r}_{i+1,i} }{ |{\bf r}_{i+1,i}|^2}
\\
 a_i &=&\frac{ ({\bf r}_i -{\bf r}) \times {\bf r}_{i+1,i} \cdot {\bf n}}
{|{\bf r}_{i+1,i}|^2}
,~~h_i  = \frac{|({\bf r}_i -{\bf r}) \cdot {\bf n}|}{|{\bf r}_{i+1,i}|}
,~~
 {\bf n}
= \frac{{\bf r}_{2,1} \times {\bf r}_{3,1}}
              {|{\bf r}_{2,1} \times {\bf r}_{3,1}|} \nonumber
\end{eqnarray*}
The remaining two integrations are done numerically using a  $N_g=3$-point  or  $N_g=7$-point 
Gauss quadrature formula for a triangle domain                

\begin{equation} \label{404}
   \int_{\Delta} df \int_{\Delta^\prime} df^\prime\
 \frac{1} {\mid {\bf r}-{\bf r}^\prime \mid}
 = \sum_{i=1}^{N_g}  w_i~ \int_{\Delta^\prime}~df^\prime \frac{1}
{\mid
 {\bf r}_1+{\bf r}_{2,1}~ \zeta_i +
{\bf r}_{3,1}~ \eta_i -{\bf r}^\prime \mid }
\end{equation}
with $\zeta_i$, $\eta_i$, and $w_i$ listed in Table I.

\vskip1.truecm
\begin{tabular}{|c|c|c|c|} \hline
$i$ & $\zeta_i$ & $\eta_i$ & $w_i $ \\ \hline
1 & 1/3 & 1/3& 270/2400 \\
2 & $(6+~\sqrt{15})/21$ & $(6+~\sqrt{15})/21$ & $(155+\sqrt{15})/2400$ \\
3 & $(9-2\sqrt{15})/21$ & $(6+~\sqrt{15})/21$ & $(155+\sqrt{15})/2400$ \\
4 & $(6+~\sqrt{15})/21$ & $(9-2\sqrt{15})/21$ & $(155+\sqrt{15})/2400$ \\
5 & $(6-~\sqrt{15})/21$ & $(6-~\sqrt{15})/21$ & $(155-\sqrt{15})/2400$ \\
6 & $(9+2\sqrt{15})/21$ & $(6-~\sqrt{15})/21$ & $(155-\sqrt{15})/2400$ \\
7 & $(6-~\sqrt{15})/21$ & $(9+2\sqrt{15})/21$ & $(155-\sqrt{15})/2400$ \\ \hline
\end{tabular}   ~~~~~~~
\begin{tabular}{|c|c|c|c|} \hline
$i$ & $\zeta_i$ & $\eta_i$ & $w_i $ \\ \hline
1 &~ 1/6~ &~ 1/6 ~&~ 1/6 ~\\
2 &~ 2/3~ &~ 1/6~ &~ 1/6 ~\\
3 &~ 1/6~ &~ 2/3~ &~ 1/6 ~\\ \hline
\end{tabular}
\vskip0.3truecm
\centerline{\hskip1.truecm (a) \hskip8.truecm (b)}
\vskip0.2truecm
\centerline{ {\bf Table I} Gauss quadrature with weights $w_i$ at points $\zeta_i$ and $\eta_i$} 
\centerline{ for a $N_g=7$ point (a) and  $N_g=3$ point (b) formula}
\vskip.5truecm

For the self-inductance of a triangle one gets
\begin{equation} \label{405}
 L_{ik} =  {\bf e}_i\cdot {\bf e}_{k} 
 \frac{|{\bf r}_{2,1} \times {\bf r}_{3,1}|^2}{12 \pi}
 \sum_{j=1}^3
\frac{1}{|{\bf r}_{j+1,j}|}
    \ln(\frac{L}{L-2~|{\bf r}_{j+1,j}|})
\end{equation}
with $L=|{\bf r}_{2,1}|+|{\bf r}_{3,2}|+|{\bf r}_{1,3}|$.
\vskip.3truecm

 The magnetic field ${\bf B}_\Delta$ produced by a  constant current ${\bf j}_\Delta$ on the triangle  can be computed
 analytically
\begin{equation} \label{406}
  {\bf B}_\Delta = - \frac{1}{4 \pi}   {\bf j}_\Delta \times \nabla \int_{\Delta
^\prime}~df^\prime \frac{1}{|{\bf r} -
{\bf r}^{\prime}|}
\end{equation}
with
\begin{eqnarray} \label{407}
\nabla  \int_{\Delta^\prime}~df^\prime \frac{1}{|{\bf r} - {\bf r}^{\prime}|} &
= &
\sum_{i=1}^3\Big (\frac{ {\bf n} \times {\bf r}_{i+1,i}}{|{\bf r}_{i+1,i}|}
\ln \frac{l_{i}^++l_{i}^-+1}{l_{i}^++l_{i}^--1} \\
& +&  {\bf n} \frac{({\bf r}_{i} - {\bf r})\cdot {\bf n}}{|({\bf r}_{i} - {\bf r
})\cdot {\bf n}|}~
\Big(\arctan \frac{a_i~ d_i^+}{a_i^2+h_i^2+ l_i^+ h_i} -
\arctan \frac{a_i~ d_i^-}{a_i^2+h_i^2+ l_i^- h_i} \Big) \Big )\nonumber 
\end{eqnarray}
 so that the contribution of a triangle to the vacuum energy term (\ref{305}) is given by
\begin{equation} \label{408}
W_{\Delta}  = \frac{1}{2 } \int_{S_1}  du~dv
(\nabla s \cdot {\bm \xi}) (F^\prime_T {\bf r}_{1,v} + F^\prime_P {\bf r}_{1,u}) \cdot {\bf B}_\Delta
\end{equation}
The surface-current potential  on
a poloidally and toroidally closed surface
consists of two multivalued secular terms determining the net-poloidal and net-toroidal
current  and of a single-valued periodic term $\phi$.
The current potential
is approximated by its value
at the numbered global nodes of the triangulated domain.   
There are three local nodes at the vertices of each triangle. The nodes are numbered
anti-clockwise.
One defines  
the vector of the nodal values at the vertices of all  triangles  by
\begin{eqnarray}   \label{409}
\bar \Phi(i) &= & U(i)~I^T +V(i)~I^P  + \sum_{j=1}^{N_{ind}}~{\bf H}(i,j)~\phi(j),~~i=1,\cdots,3 N_{t}                         
\end{eqnarray}
where $N_t$  is the number of triangles 
and $N_{ind}$ is the number of independent variables $\phi(j)$ at the global nodes.
A global node belongs to several triangles.
$U(i)$ and $V(i)$ are the nodal values of a unit net-toroidal and net-poloidal current distribution  
which can be arbitrarely chosen.
The so-called connectivity matrix ${\bf H}$ relates the global nodes to the local nodes of the  triangles.
If there are holes in the wall the current potential has to satisfy the
boundary condition $\Phi = const.$ along the edges: the nodes at
the vertices of the triangles along the edges of a hole have the same value 
reducing 
the number of the independent variables.
 That has to be incorporated in the connectivity matrix.

The matrix elements of a Lagrangian ${\cal L}_S = \frac{1}{2} \bar \Phi^\top {\bf \overline{M}} \bar \Phi$ are
\begin{equation}  \label{410}
\overline{{\bf  M}}_{3i-3+k,~3 i^\prime -3 +k^\prime} = \frac{{\bf e}^i_k \cdot {\bf e}^{i^\prime}_{k^\prime} ~}{4 \pi} 
  \! \int_{{\Delta}^\prime}  df'\! \int_{\Delta} df\
 \frac{1} {\mid {\bf r}_{i^\prime}-{\bf r}_i \mid},
\left\{ \begin{array} {r@{\quad}l}
         k=1,2,3,~~i=1,\cdots, N_t \\
         k^\prime =1,2,3,~~i^\prime  =1,\cdots , N_t 
         \end{array}  \right.
\end{equation}
The contribution of a triangulated (resistive) wall to the Lagrangian 
is then given by
\begin{equation} \label{411}
{\cal L}_S = \frac{1}{2} \Phi^\top ~ {\bf M}_{tt}~\Phi, ~ {\bf M}_{tt}= {\bf G}^\top {\bf \overline{M} }{\bf G},~~{\bf G}  = (U,V,{\bf H}),~~ 
\Phi =\left( \begin{array}{c} I^T  \\  I^P \\ \phi
\end{array} \right),
\end{equation}
where $\Phi$ is  the vector of the current potential  of the independent variables.
The contribution to the vacuum energy matrix  of the wall  is obtained with
(\ref{406}),(\ref{407}),(\ref{410}) and (\ref{411}).

\begin{equation}
 {\bf M}_{\hat \xi t}  = 
 \overline{{\bf M}}_{\hat \xi \overline{\Phi}} {\bf G},
\end{equation}
where the matrix    elements of~ 
$ \overline{{\bf M}}_{\hat \xi \overline{\Phi}}$ 
are the contributions to the Fourier harmonics of $ \nabla s  \cdot {\bm \xi}$
for the nodal values of the vertices of all triangles.

One possible choice  for the net-current contribution for a closed torus without holes can be defined as follows:
On a rectangular mesh  with $n_u$ poloidal and $n_v$ toroidal meshpoints
\begin{eqnarray}
(u_i,v_k)= \left(\frac{i-1}{n_u},\frac{k-1}{n_v}\right ), {{i=1,\cdots,n_u+1}\atop
                                               {k=1,\cdots,n_v+1 }} 
\end{eqnarray}
the positions of the $2 n_u n_v$ triangles are given by
\begin{eqnarray}
\bar {\bf r}_{l-1,1}\!&=&\!{\bf r}(u_{i}~~~,v_{k+1}),~
\bar {\bf r}_{l,1}  = {\bf r}(u_{i+1},v_{k}~~~),   \\
\bar {\bf r}_{l-1,2}\!&=&\!{\bf r}(u_{i+1},v_{k+1}),~
\bar {\bf r}_{l,2}  = {\bf r}(u_{i}~~~,v_{k}~~~),
~~l=2(i+n_u(k\!-\!1)),
   {i= 1,n_u\! \atop
   k = 1,n_v\!}
   \nonumber   \\
\bar {\bf r}_{l-1,3}\!&=&\!{\bf r}(u_{i}~~~,v_{k}~~~),~
\bar {\bf r}_{l,3}  = {\bf r}(u_{i+1},v_{k+1})
  \nonumber
\end{eqnarray}
where $l$ labels the triangles.
There are $(n_u+1)(n_v+1)$ nodes for the secular terms
being proportional to $I^T$ and $I^P$ . The
values of these current potential terms at the nodes can be chosen as  
\begin{eqnarray}
 { {U(j)= u_i}\atop{V(j)=v_k}},~~ j=i+ (n_u+1)~(k-1),~~{{i=1,\cdots,n_u+1}\atop{k=1,\cdots,n_v+1}}
\end{eqnarray}
while there are  $n_u n_v$ nodes for the periodic current potential $\phi$.

A resistive wall closed only poloidally or toroidally is topologically  a cylinder.
In that case the current potential becomes single-valued. For a poloidally (toroidally)
closed wall the net-toroidal (net-poloidal) current vanishes and the net-poloidal
(net-toroidal) current  is given by the difference of the current potential value at 
the two boundaries of the wall.

%% file: kapitel_cor.tex
\par
\medskip
Numerical results are presented for an ASDEX Upgrade-type test 
equilibrium which
is unstable with respect to external modes. 
The plasma equilibrium properties are: major radius $R_0=1.64$~m,
plasma current $I_p=0.98$ MA, monotone $q$-profile with $q_{axis}=1.46$
and $q_{boundary}=5.26$, vacuum magnetic field strength $B_0(R_0)=2.43$~T,
and volume average beta $<\beta>= 2.58\%$. A poloidal cross-section
of the flux surfaces, as well as the  pressure and  $q-$profiles
are shown in Figs~1a-c. In Fig.~1a the positions of the plasma boundary 
(red), the
resistive wall (blue), and the toroidal field coils (brown) are sketched. 
At the low field side
the plasma boundary extends to $R\approx 2.14$~m, while the resistive wall
is localized at $R \approx 2.23$~m, so that  the plasma-wall distance
amounts to $\Delta R \approx 9$~cm at this position.
\par
\medskip
\begin{minipage}{7.5truecm}
\centerline{
\includegraphics*[scale=0.6]
     {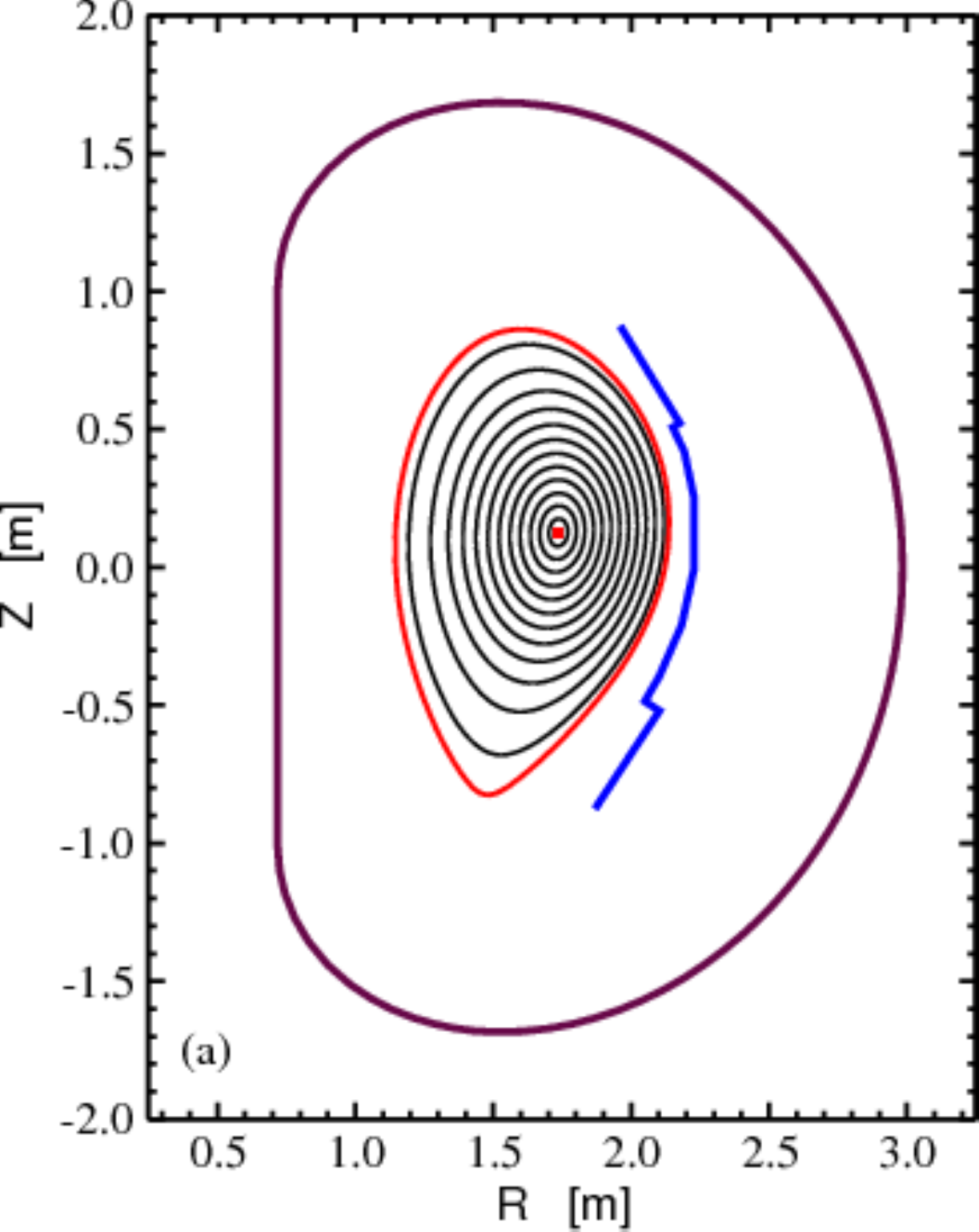}
}
\vskip -3.25 truecm
\centerline{
 \large{~~~~~~~~~~~~~~~~~ $\color{red}{\bf S}_1  ~~~ \color{blue}{\bf S}_2 ~~~ ~~~~~~~\color{brown}{\bf  S}_3$}}
\end{minipage}
\hskip 0.8 truecm
\begin{minipage}{7.5truecm}
\centerline{
\includegraphics*[scale=0.5]
    {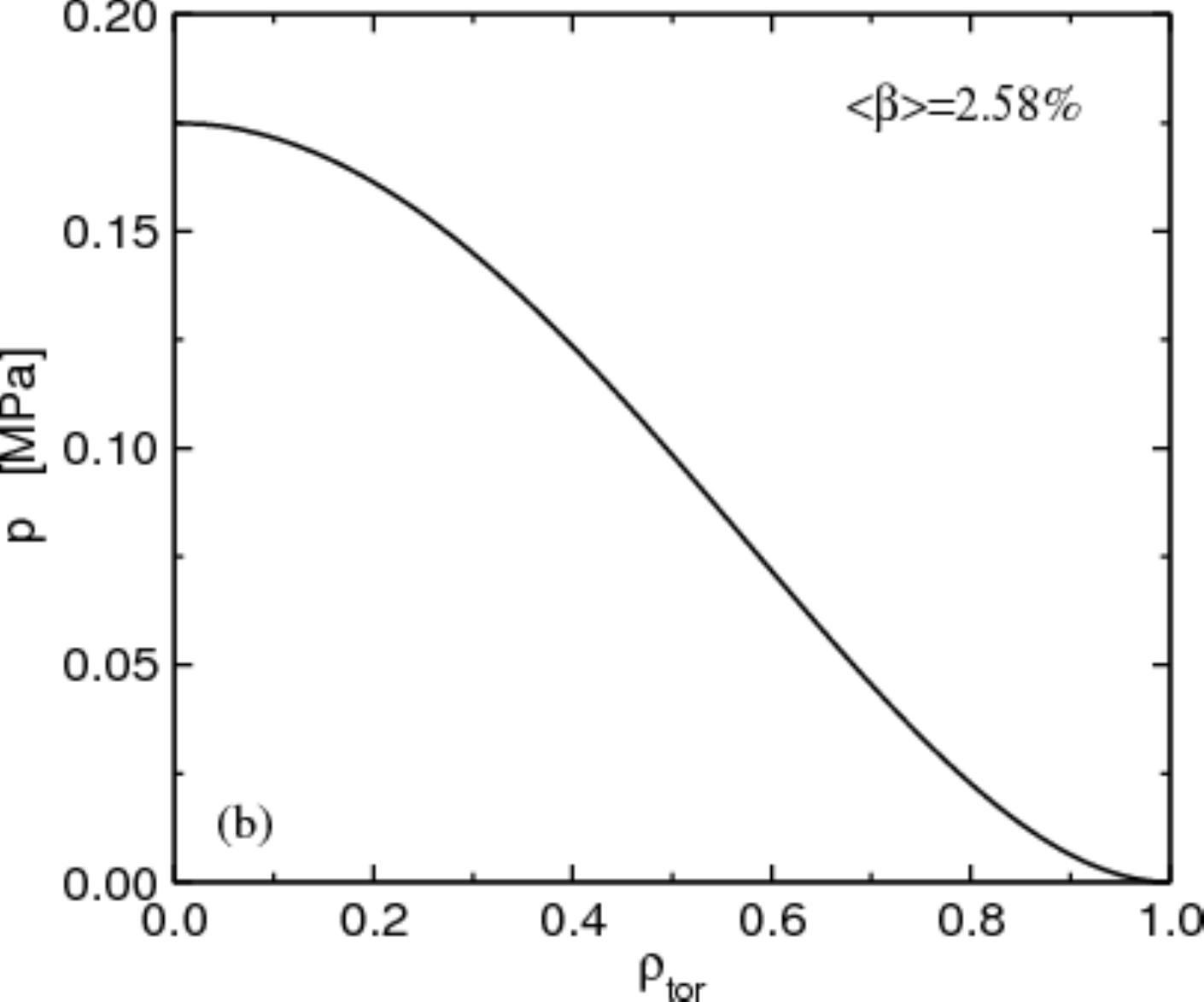}
}
\par
\medskip
\centerline{
\hskip 0.5truecm
\includegraphics*[scale=0.5]
    {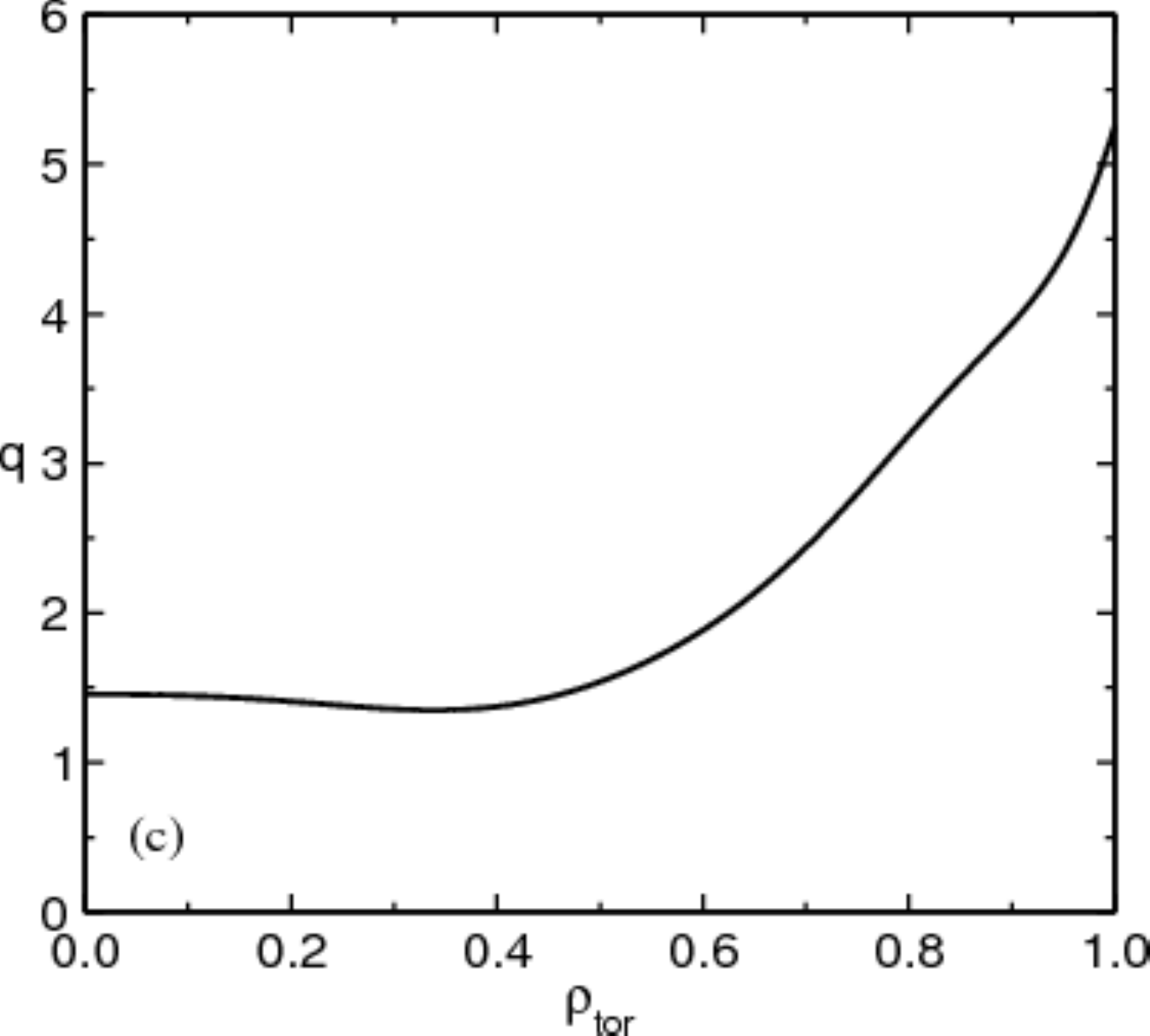}
}
\end{minipage}
\par
{\bf Fig.~1:} 
{\sl (a) Poloidal cross-section of flux surfaces, plasma boundary (S$_1$),
resistive wall (S$_2$), and toroidal field coil (S$_3$),
(b) pressure profile,
and (c) $q$-profile of the ASDEX Upgrade-type test equilibrium 
($\rho_{tor} = \sqrt{s}$).}
\par
\bigskip
Figure 2a shows a 3D view of the preliminary wall design of ASDEX Upgrade
\cite{Suttrop2009}. In Fig.~2b the whole configuration composed of
plasma, resistive wall, and 16 toroidal field coils is presented. The latter
are modeled by infinitely thin bands of finite width.
\par
\bigskip
\begin{minipage}{7.5truecm}
\centerline{
\includegraphics*[scale=0.5]
    {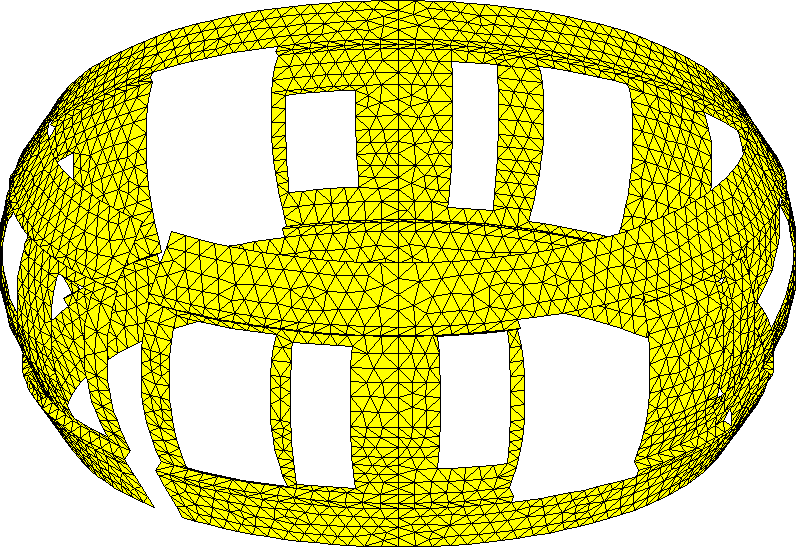}
}
\end{minipage}
\hskip 1truecm
\begin{minipage}{7.5truecm}
\centerline{
\includegraphics*[scale=0.55]
{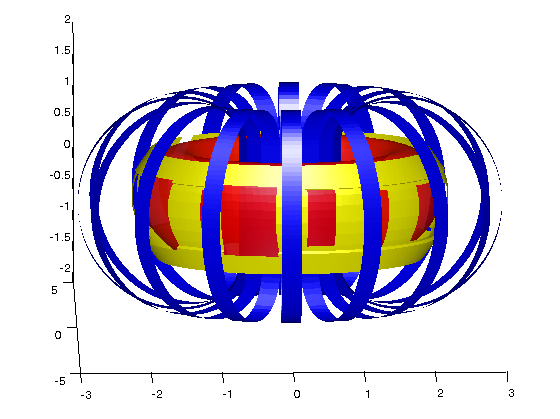}
}
\end{minipage}
\par
{\bf (a)}
\hskip  8.0truecm
{\bf (b)}
\par
{\bf Fig.~2:} 
{\sl (a) Preliminary design of the resistive wall, and (b) plasma with 
resistive wall and toroidal field coils.}
\par
\bigskip
Without wall the equilibrium is unstable with respect to $n=0,~1$ and $2$ modes
($n > 2$ are not considered). However, it can be stabilized with an ideal
conducting wall sufficiently close to the plasma (see Fig.1a)
\par
In case of a finite wall conductivity
the plasma is unstable on a resistive time scale. A surface conductivity
$\sigma d = 2.8\cdot 10^5$~S was used with $\sigma$ being the specific
conductivity and $d$ the thickness of the wall.
\par
In Fig.~3a the eigenvalue $\gamma$ of the $n=0$ mode is plotted versus the
plasma-wall distance
$r_{shift}$. The eigenvalue is computed with the CAS3D code including
the perturbed kinetic energy term. The $m=0,n=0$ harmonic causes
a current in the toroidal field coils producing a
net-toroidal field in the region between the plasma boundary and
the coils.
In Fig.~4 the $m$-harmonics of the $n=0$ eigenfunction are plotted for the
no-wall limit case ($\gamma=510481$ 1/s). There,
the $m=0$ harmonic only makes a very small contribution.
\par
\bigskip
\begin{minipage}{7.50truecm}
\centerline{
\includegraphics*[scale=0.45]
    {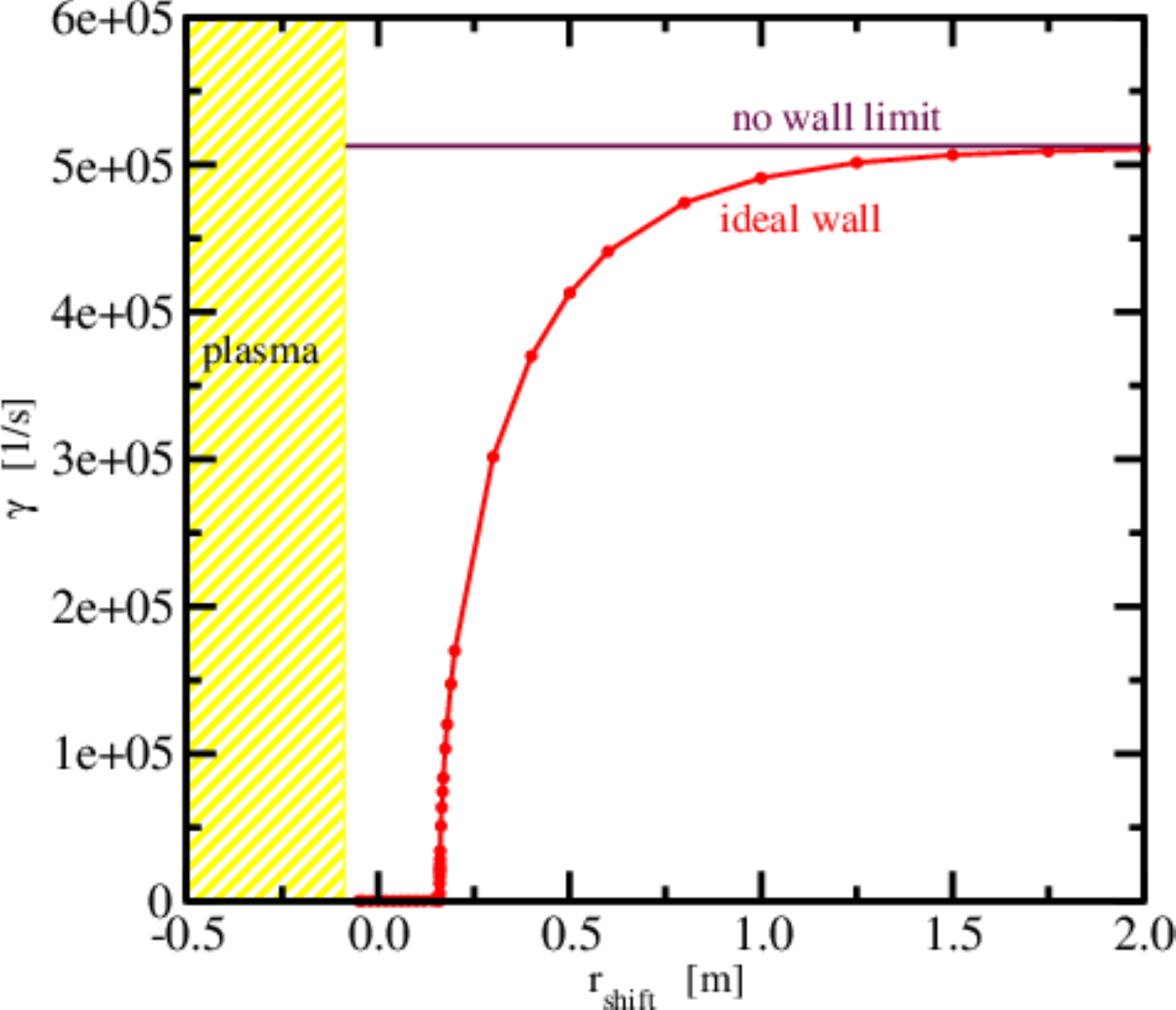}
}
\end{minipage}
\hskip 1.0truecm
\begin{minipage}{7.50truecm}
\centerline{
\includegraphics*[scale=0.45]
    {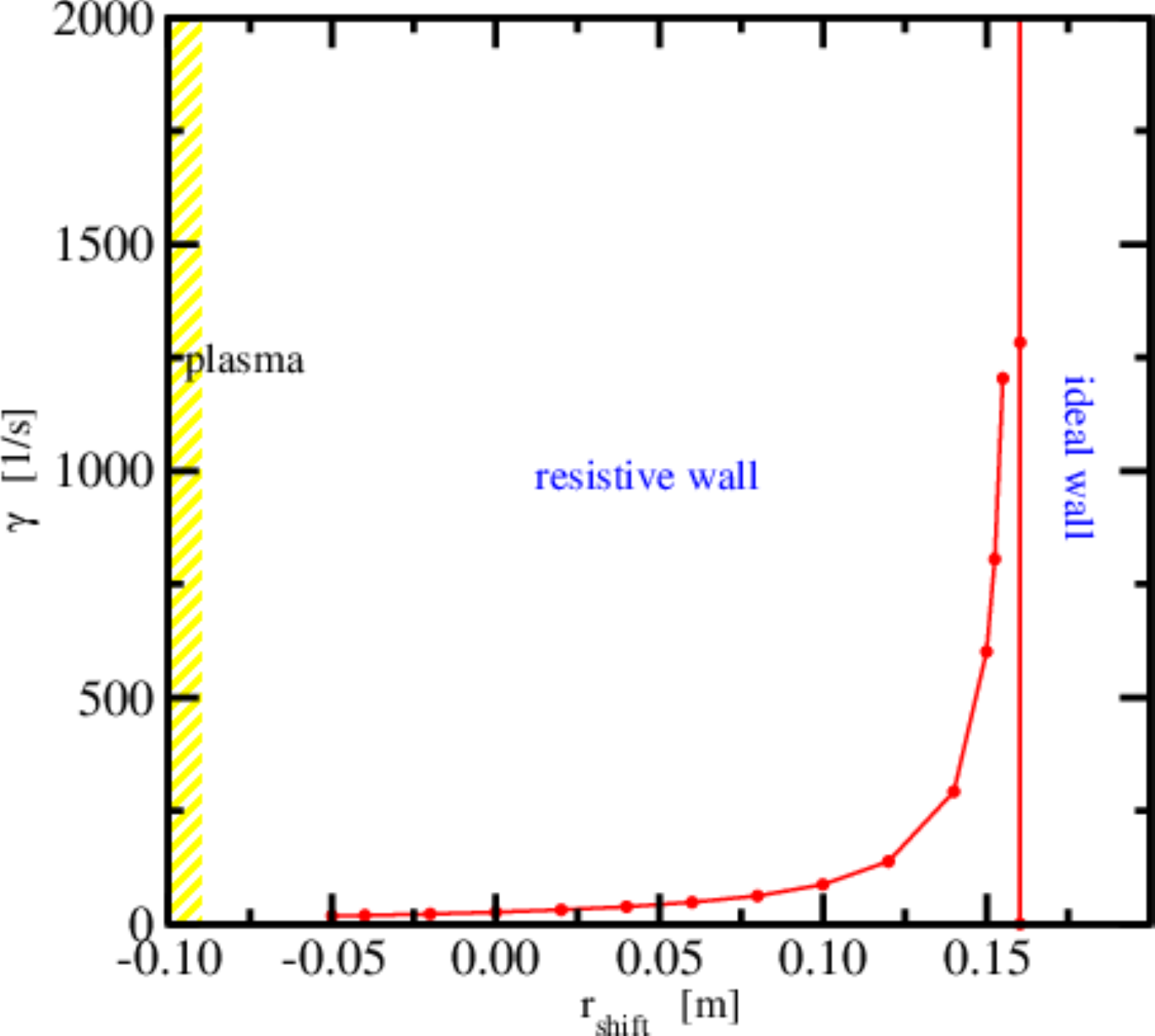}
}
\end{minipage}
\vskip -.5truecm
{\bf (a)}
\hskip  8.truecm
{\bf (b)}
\par
{\bf Fig.~3:} 
{\sl Growth rate of the n=0 mode in dependence of the wall position. The
latter is presented by the radial shift of the wall with $r_{shift}=0$
being the planned position of the wall in ASDEX Upgrade. The hatched 
area marks the
plasma with its boundary lying approximately 9~cm inside the unshifted wall.
The plots show the growth rates assuming an ideal, and a resisitve
wall, respectively. The plot on the right-hand side presents an enlargement
of the left-hand side plot, showing the resistive wall growth rates and
the stabilizing ideal wall position at $r_{shift}=0.16$~m in detail.}
\par
\bigskip  
\medskip
\centerline{
\includegraphics*[scale=0.55]
    {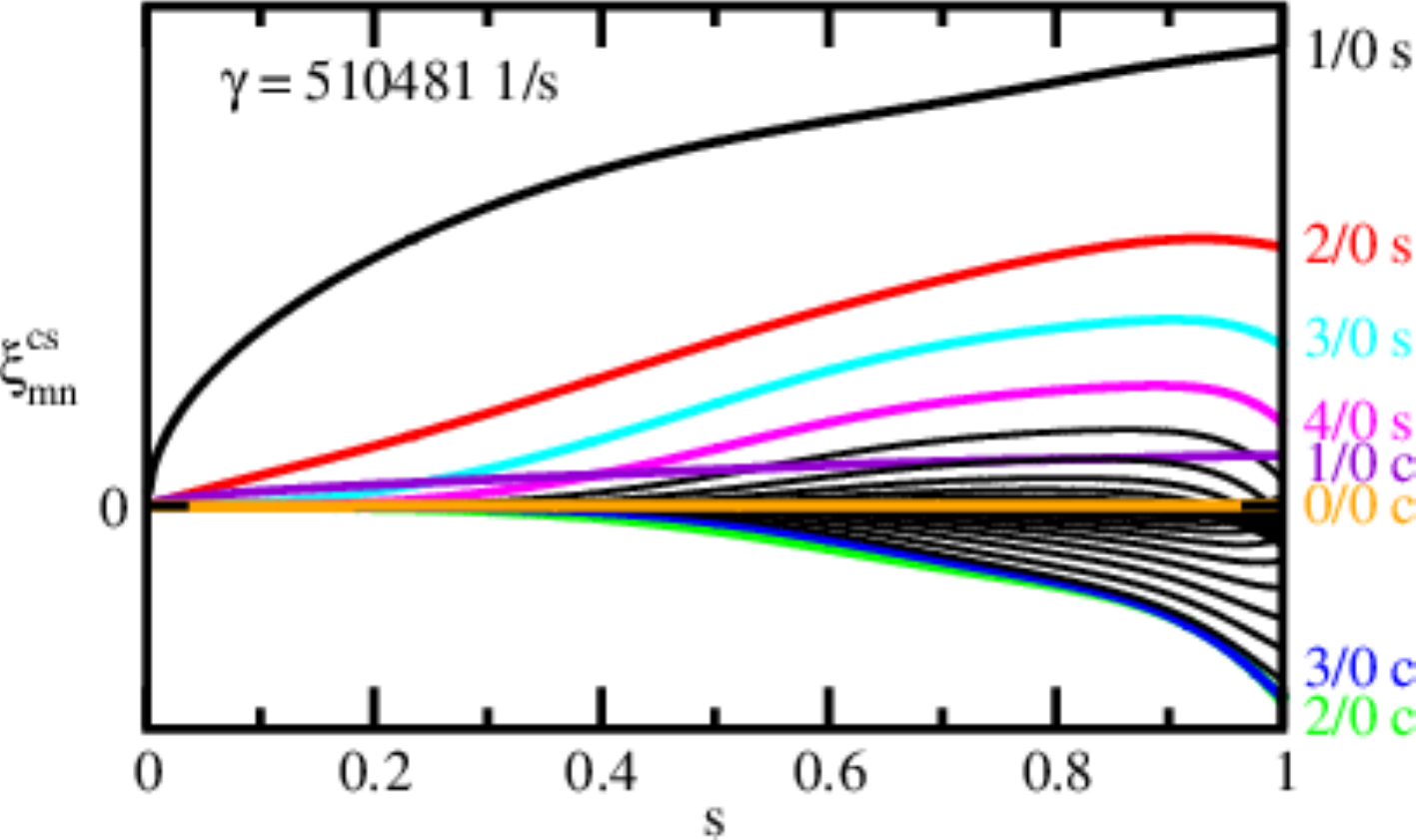}
}
{\bf Fig.~4:}
{\sl Eigenfunction of the $n=0$ mode represented by the
cosine and sine Fourier harmonics, $\xi^{c}_{mn}, \xi^s_{mn}$,
of the normal component
of the displacement vector $\bm \xi$ as function of $s$.
Here and in the following plots, the
largest harmonics are marked by their poloidal and toroidal indices, $m/n$.
The s and c attached to these numbers characterize the sine and
cosine harmonics, respectively. In this plot, additionally, the m=0, n=0
harmonic is marked by the orange line.}
\vfill
\eject
\par
\bigskip
\begin{minipage}{7.50truecm}
\centerline{\small {n=0, $\gamma=26$ 1/s} \normalsize}
\par
\includegraphics*[scale=0.475]
    {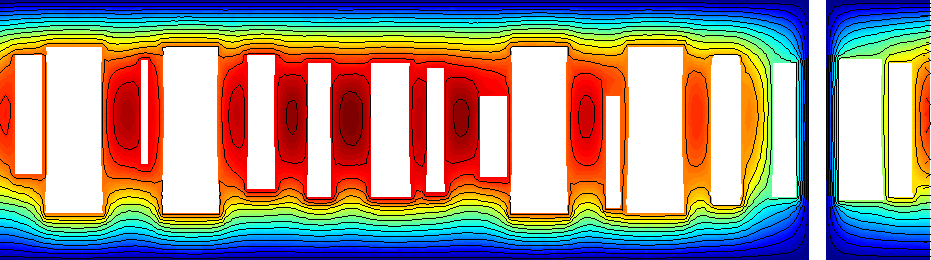}
\par
\medskip
\centerline{\small {n=1, $\gamma=67$ 1/s} \normalsize}
\par
\includegraphics*[scale=0.475]
    {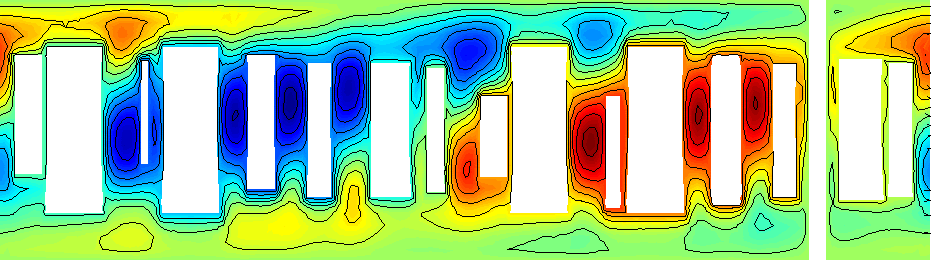}
\par
\medskip
\centerline{\small {n=1, $\gamma=50$ 1/s} \normalsize}
\par
\includegraphics*[scale=0.475]
    {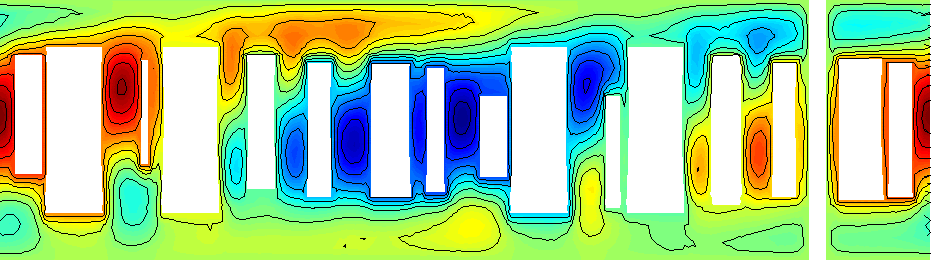}
\par
\medskip
\centerline{\small {n=2, $\gamma=314$ 1/s} \normalsize}
\par
\includegraphics*[scale=0.475]
    {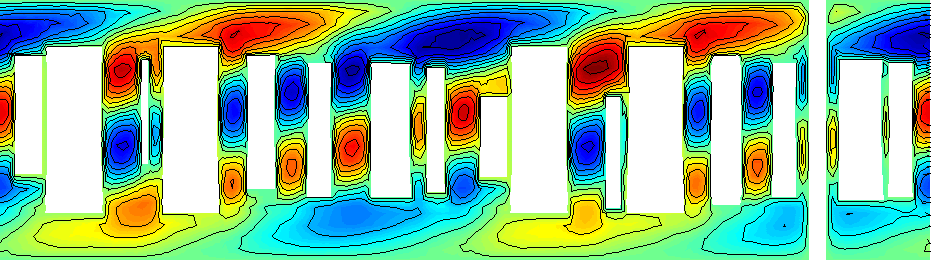}
\par
\medskip
\centerline{\small {n=2, $\gamma=276$ 1/s} \normalsize}
\par
\includegraphics*[scale=0.475]
    {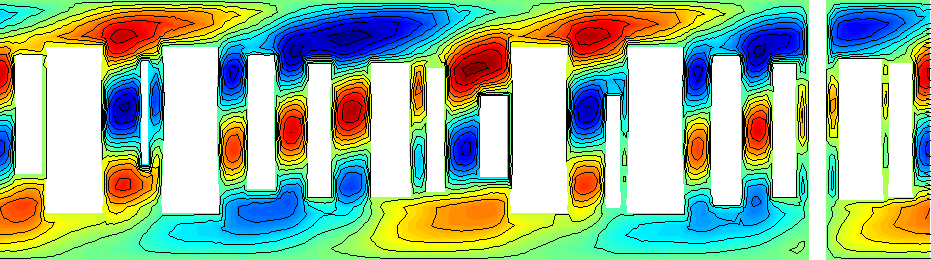}
\end{minipage}
\hskip 1.0truecm
\begin{minipage}{7.50truecm}
\centerline{\small {n=0,1,2, $\gamma=24$ 1/s} \normalsize}
\par
\includegraphics*[scale=0.475]
    {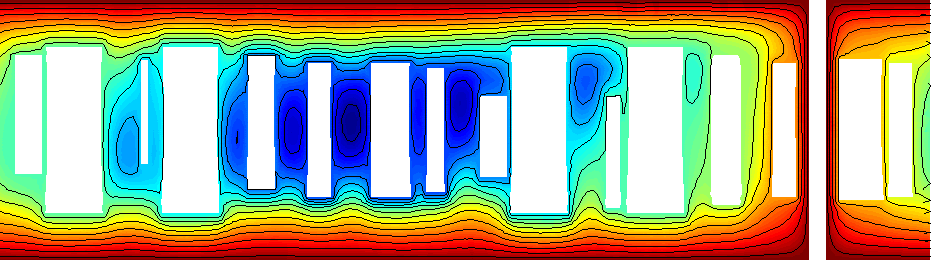}
\par
\medskip
\centerline{\small {n=0,1,2, $\gamma=67$ 1/s} \normalsize}
\par
\includegraphics*[scale=0.475]
    {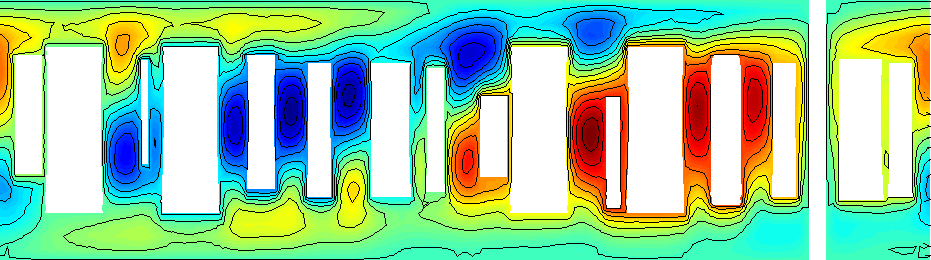}
\par
\medskip
\centerline{\small {n=0,1,2, $\gamma=51$ 1/s} \normalsize}
\par
\includegraphics*[scale=0.475]
    {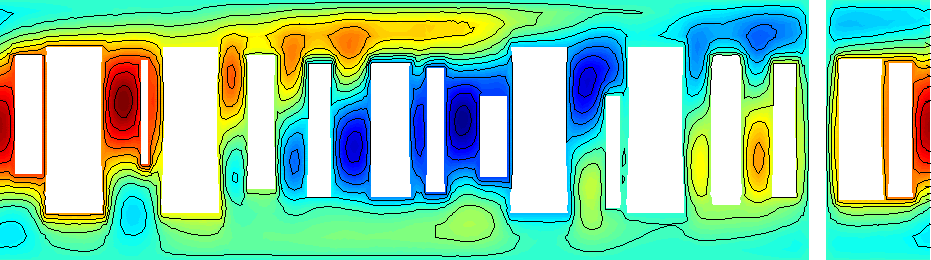}
\par
\medskip
\centerline{\small {n=0,1,2, $\gamma=386$ 1/s} \normalsize}
\par
\includegraphics*[scale=0.475]
    {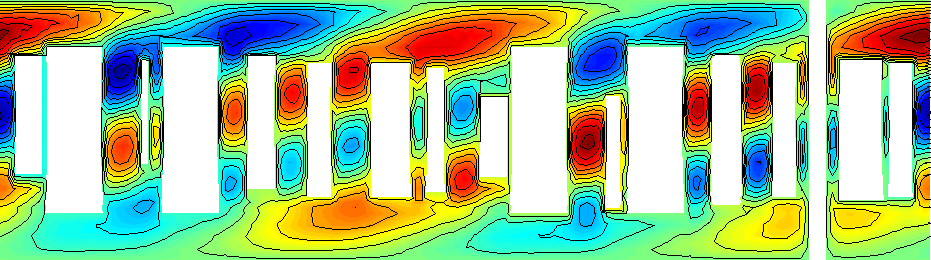}
\par
\medskip
\centerline{\small {n=0,1,2, $\gamma=346$ 1/s} \normalsize}
\par
\includegraphics*[scale=0.475]
    {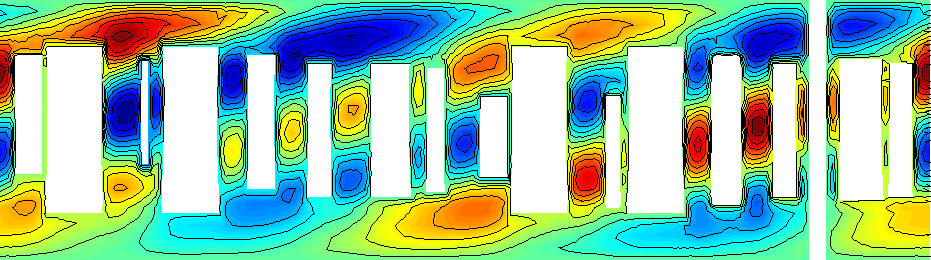}
\end{minipage}
\par
{\bf (a)}
\hskip 8.truecm
{\bf (b)}
\par
{\bf Fig.~5:} 
{\sl Resistive wall surface currents belonging to the 5 unstable $(n=0,1,2)$
multi-mode eigenvalues (b) and to the unstable single $n=0$, $n=1$, $n=2$ 
eigenvalues (a) are shown.
The current pattern of the multi-mode cases  can be uniqely related
to the single $n$ solutions.}
\par
\bigskip
The axisymmetry is broken by the multiply-connected ASDEX Upgrade
resistive wall
so that, in principle, all
$n$-harmonics  contribute to an eigenmode.
In Fig.~5b surface-current lines of the induced wall  currents are shown
for the 5 unstable eigenmodes where $n=0,1,2$ toroidal harmonics
have been taken into account.
For comparison, in Fig.~5a surface current are shown obtained  from single
 $n=0$, $n=1$ and $n=2$ toroidal mode computation.
Comparing the eigenvalues - also quoted in Figs~5a-b - one can uniquely
relate the single $n$ modes to the $n=0,1,2$ eigenmodes.
That is, each eigenmode is dominated by one $n$-harmonic.
 The asymmetry of the wall geometry leads to a significant splitting of the
eigenvalues and coupling of the toroidal harmonics.
In Figs~6a-b the two orthogonal eigenfunctions for the $(n=2)$ case with
different eigenvalues (see Fig.~5a row 4 and 5)  are shown.
In order to get a sufficiently good resolution
 in the neighbourhood of rational surfaces a non-equidistant radial
grid has been implemented
in the CAS3D code.
Figure~7a shows the coupling of the toroidal $n=1,n=2$ harmonics. In Fig.~7b
the enlargement of the 3/2-harmonic region demonstrates the improved
resolution obtained by the
accumulated mesh at the $q=3/2$ surface.
\par
\bigskip
\begin{minipage}{7.50truecm}
\centerline{
\includegraphics*[scale=0.55]
    {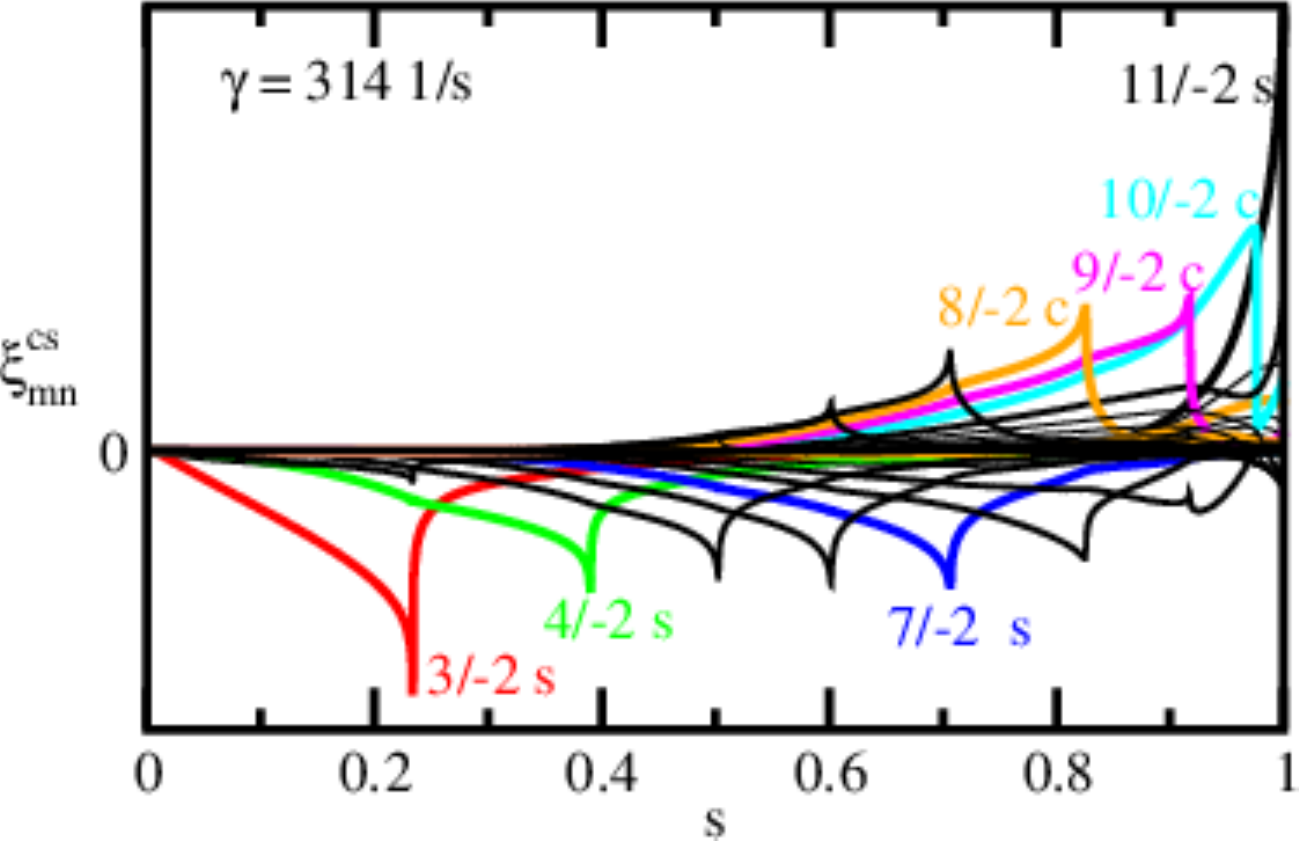}
}
\end{minipage}
\hskip 1.0truecm
\begin{minipage}{7.50truecm}
\centerline{
\includegraphics*[scale=0.55]
    {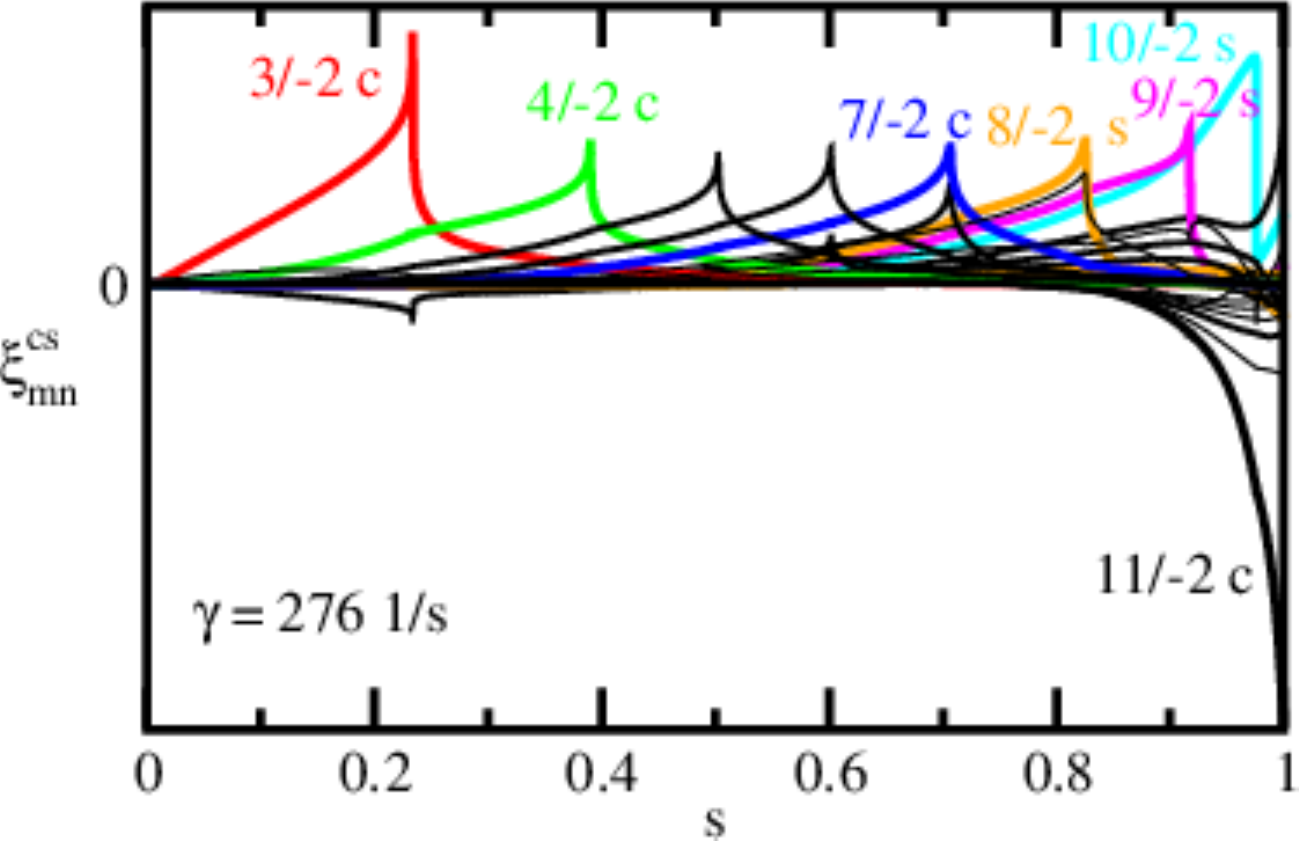}
}
\end{minipage}
\par
{\bf (a)}
\hskip  8.0truecm
{\bf (b)}
\par
{\bf Fig.~6:}
{\sl Eigenfunctions of the $n=2$ mode in presence of a resistive wall
and ideal toroidal field coils. The two plots show the eigenfunction
spectra for the two eigenvalues which correspond to the two
existing orthogonal solutions.}
\par
\bigskip
\begin{minipage}{7.50truecm}
\centerline{
\includegraphics*[scale=0.55]
    {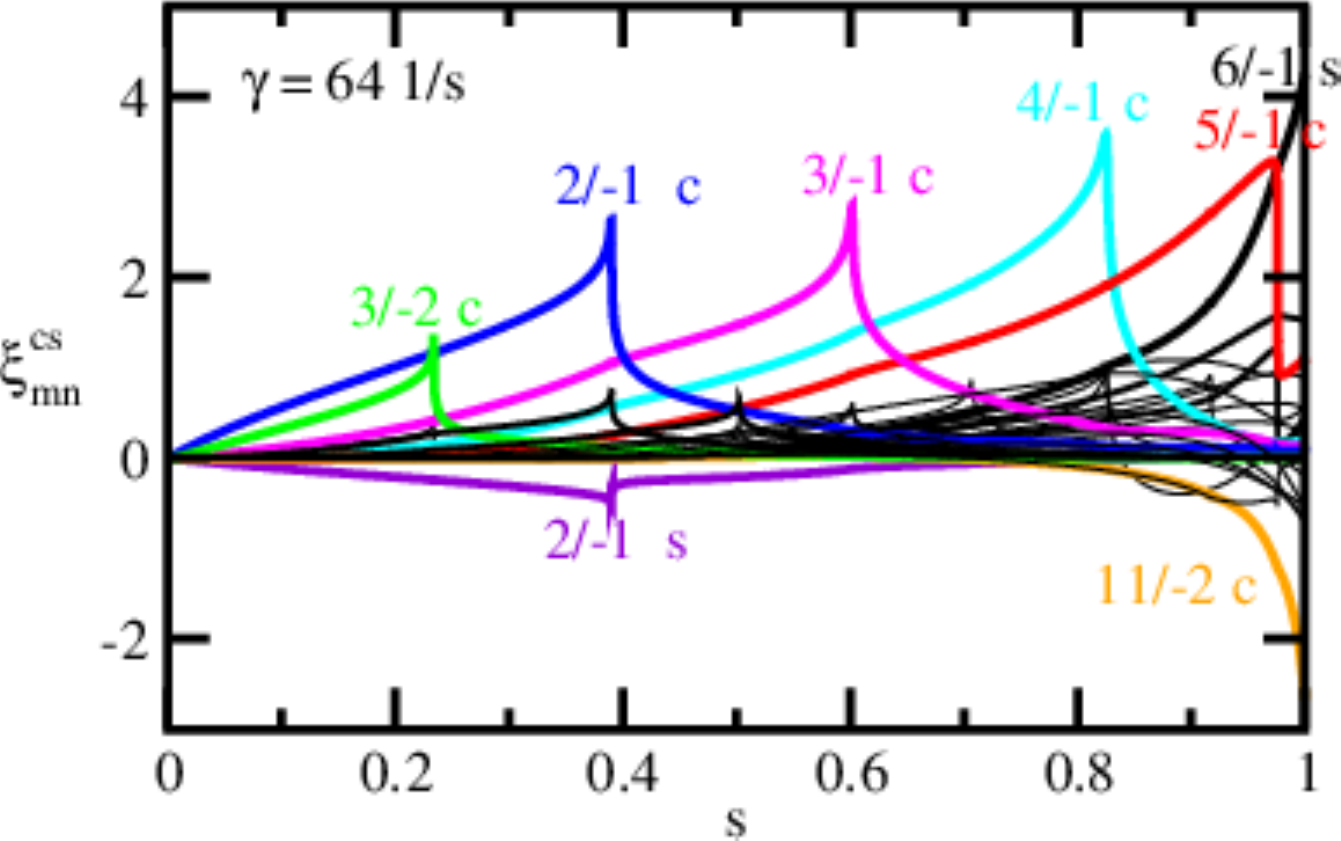}
}
\end{minipage}
\hskip 1.0truecm
\begin{minipage}{7.50truecm}
\centerline{
\includegraphics*[scale=0.55]
    {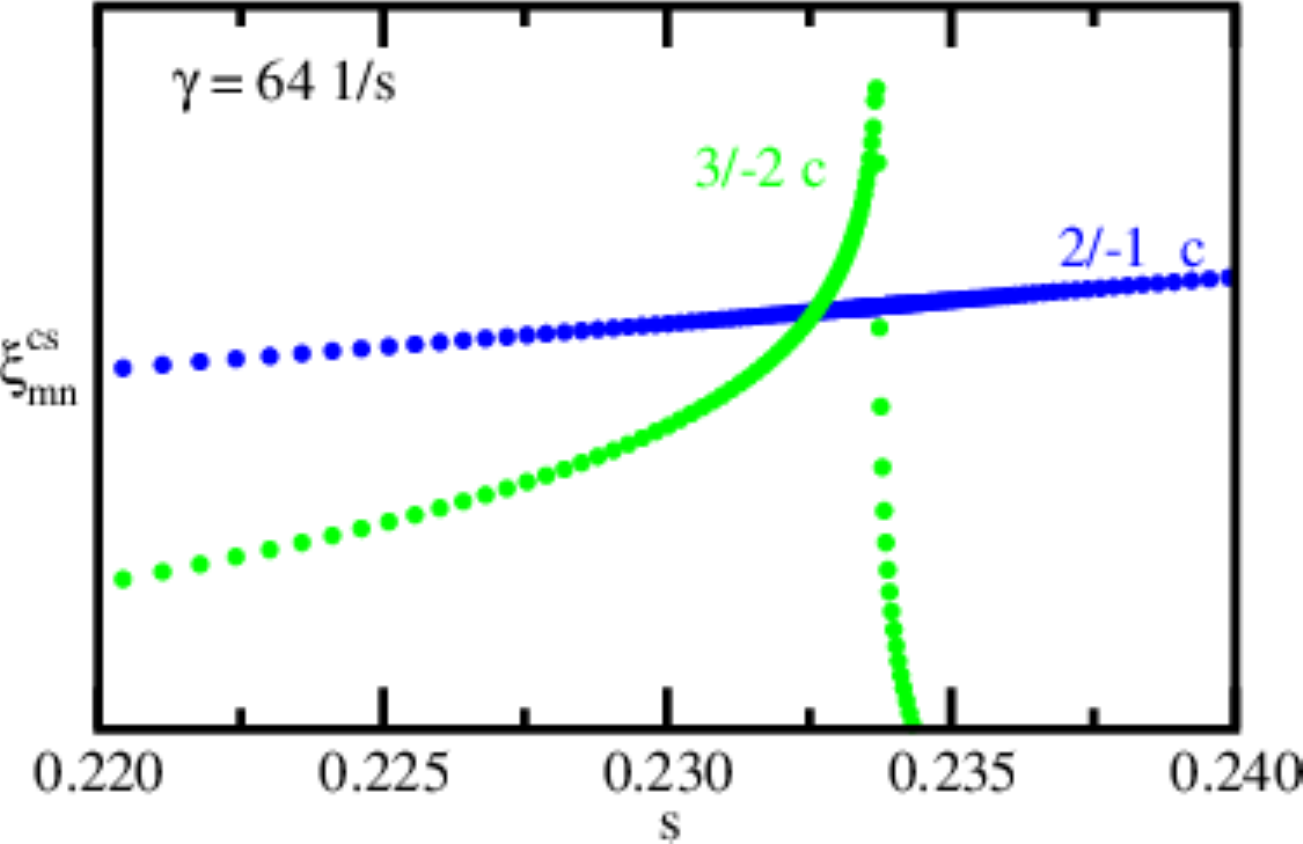}
}
\end{minipage}
\par
{\bf (a)}
\hskip  8.0truecm
{\bf (b)}
\par
{\bf Fig.~7:}
{\sl (a) Eigenfunctions of the coupled $n=1$, $n=2$ mode. (b)
Enlargement of the 3/2-harmonic showing the radial grid accumulation
around the rational $q=3/2$ surface. To obtain the same
grid refinement with an equidistant grid 40000 grid points would be
necessary compared  to 931 non-equidistant grid points.}
\par
\bigskip
\bigskip

%% file: outlook_m.tex
\section{Outlook}
\par
\medskip
The STARWALL code is a very flexible numerical tool which can be
combined with linear and nonlinear stability codes. {The non-linear MHD code
JOREK and the linear MHD code CASTOR3D are currently under development using STARWALL
for the resistive wall part.}

{The JOREK-STARWALL code has already been applied to
nonlinear studies of resistive wall
modes~\cite{McAdams2014}, vertical displacement events~\cite{Hoelzl2014}, disruptions triggered
by massive gas injection~\cite{Fil2015} and QH-Mode~\cite{Liu2014}. Besides a current optimization
and parallelization of STARWALL for resolving larger problem sizes, an extension of
JOREK-STARWALL to include Halo currents
is currently under development~\cite{Hoelzl2014}. 

{Furthermore, the STARWALL code is used in combination with the CASTOR$\_$3DW
code for stability studies of axisymmetric equilibria in presence of ideal wall structures,
including physical effects such as plasma resistivity, viscosity, poloidal and
toroidal rotation~\cite{Strumberger2011}. Currently, CASTOR3D is under
development~\cite{Strumberger2014a}, a hybrid of the CASTOR$\_$3DW and the STARWALL codes.
Solving an extended eigenvalue problem (MHD equations of CASTOR\_3DW and vacuum part of STARWALL),
this code takes plasma inertia and resistive walls simultaneously into account. Besides
the straight field line coordinates used for the plasma part so far, general flux
coordinates are implemented such that linear studies of resistive and rotating 3D
equilibria in presence of 3D resistive walls will be possible.}
\par
\bigskip

%% file: lagrangian.tex
\par 
\medskip
\section{Lagrangians}
\label{sec:lagrangian}
The variational method will be explained  by starting with a simple problem.
For a current ${\bf j}$ on an ideal conducting torus one can define the following  variational
problem: determine the current distribution by minimizing the energy of the magnetic field  produced. The Lagrangian - being the magnetic energy - is given by
\begin{equation} \label{100a}
{\cal L}_{S}  = \!\frac{1}{8\pi} \int \limits_{S} \! df \int \limits_{S} \!df^\prime
\frac{{\bf j} \cdot {\bf j}^\prime} {\mid {\bf r}-{\bf r}^\prime \mid}
\end{equation}
with the divergence-free surface current
\begin{equation}
  {\bf j} = {\bf n} \times \nabla \Phi,~~\Phi = I^P~v + I^T~u + \phi(u,v)
\end{equation}
with angle-like surface coordinates $u,v: 0 \le u \le 1,0 \le v \le 1$.

 Given a net-poloidal current $ I^P$ and/or a net-toroidal current $I^T$
one gets the "Euler equation" by varying ${\cal L}_{S}$ with respect to the single-valued
potential $\phi(u,v)$.
\begin{equation}
\delta {\cal L}_{S}  = \!\frac{1}{4\pi} \int \limits_{S} \! df \int \limits_{S} \!df^\prime
\frac{\delta {\bf j} \cdot {\bf j}^\prime} {\mid {\bf r}-{\bf r}^\prime \mid}
 = \int \limits_{S} \! df~ \delta {\bf j} \cdot {\bf A}
\end{equation}
and with 
\begin{equation}
      \delta {\bf j} = \frac{ {\bf r}_u \frac{\partial}{\partial v} \delta \phi
      -{\bf r}_v \frac{\partial}{\partial u}\delta  \phi}
            {\mid {\bf r}_v \times {\bf r}_u \mid}
\end{equation}
\begin{eqnarray}
\delta {\cal L}_{S}  &= &\int \limits_{S} \! du~dv 
      ( {\bf r}_u \frac{\partial}{\partial v} \delta \phi
      -{\bf r}_v \frac{\partial}{\partial u}\delta  \phi) \cdot {\bf A} 
                       = -\int \limits_{S} \! du~dv  \delta \phi
      ( {\bf r}_u \frac{\partial}{\partial v} 
      -{\bf r}_v \frac{\partial}{\partial u}) \cdot {\bf A}  \nonumber \\ 
                      & = &-\int \limits_{S} \! du~dv  \delta \phi~
      [ {\bf r}_u ({\bf r}_v \cdot \nabla)
      -{\bf r}_v ({\bf r}_u \cdot \nabla)] \cdot {\bf A}       \\ 
                       &= &-\int \limits_{S} \! du~dv  \delta \phi~
      ( {\bf r}_v \times {\bf r}_u) \cdot (\nabla \times  {\bf A}) 
                       = -\int \limits_{S} \! du~dv  \delta \phi~
      ( {\bf r}_v \times {\bf r}_u) \cdot  {\bf B} \nonumber 
\end{eqnarray}
From ${\delta \cal L}_{S} = 0$ it follows  the so-called natural boundary condition
  ${\bf n} \cdot {\bf B}= 0$.

There are two independent solutions. Given a net-poloidal current $I^P$ the magnetic
field ${\bf B}$ vanishes in the exterior region of the toroidal surface and
 given a net-toroidal current $I^T$  the field ${\bf B}$ vanishes in the interior region.                 
For the problem considered one got the natural boundary condition $ {\bf n}\cdot {\bf B}=0$. In order to obtain more general boundary conditions one has to add
 appropriate terms. 
 One gets the Lagrangian ${\cal L}_V$ for the configuration treated in section 3 by terms of
type (\ref{100a}) and adding  a term linear in ${\bf j}$ appropriate  to satisfy the boundary condition at the plasma-vacuum interface $S_1$ and a  term to fulfil the boundary condition
at the resistive wall $S_2$.
\begin{equation}  \label{101a}
{\cal L}_V\!  = \!\frac{1}{8\pi} \sum_{i=1}^3 \sum_{k=1}^3 \int \limits_{S_i} \! df_i \int \limits_{S_k} \!df_k
\frac{{\bf j}_i \cdot {\bf j}_k} {\mid {\bf r}_i-{\bf r}_k \mid}
+ \!\frac{1}{2 \gamma}  \int \limits_{S_2} \!df_2 \
\frac{{\bf j}_2 \cdot {\bf j}_2 }{\sigma d} 
+\int \limits_{S_1} \!df_1~({\bf n}_1\! \cdot \! {\bm {\xi}})~ {\bf n}_1\! \cdot \! ({\bf j}_1 \times {\bf B}_0)  
\end{equation}
To study feedback stabilisation one has to extend the Lagrangian by adding terms for
sensor  and feedback coils and  for feedback voltages given by
\begin{eqnarray}  \label{102a}
 {\cal L}_S 
 &=&  \frac{1}{8\pi} \sum_{l,l^\prime=1}^{n_c,n_c} I^c_l I^c_{l^\prime} \int \limits_{C_l} \!df_l \int \limits_{C_{l^\prime}} \! df_{l^\prime} \
\frac{{\bf v}_l \cdot {\bf v}_{l^\prime} }{\mid {\bf r}_l-{\bf r}_{l^\prime} \mid}
 + \frac{1}{4\pi} \sum_{l}^{n_c} I^c_l \! \int \limits_{C_l} \! df_l 
\sum_{i=1}^3\int \limits_{S_i} \! df_{i} \
\frac{{\bf j}_i \cdot {\bf v}_l }{\mid {\bf r}_i-{\bf r}_{l} \mid}  \\
& + &\frac{1}{2 \gamma} \sum_{l=1}^{n_c} R_l {I^c_l}^2 \int \limits_{C_l}df_{l} {\bf v}_l^2
 +  \frac{1}{\gamma}\sum_{l=1}^{n_c} I^c_l~ U_{c,l}^{feedb}. \nonumber
\end{eqnarray}

with  $n_c=$ number of coils, $I^c_\ell=$ coil currents, $R_\ell=$coil resistance and feedback voltages $U^{feedb}_{c,\ell}$.

 Considering  the variation of the Lagrangian ${\cal L}_V$ (\ref {101a}) with respect to 
$\Phi_i,~i=1,2,3$  gives   

\begin{eqnarray} \label{103a}
{\delta \cal L}_V\! & =&\!\! \sum_{i=1}^3 \Big (\delta I^P_i \!  \int \limits_{S_i}\! du~dv   
~{\bf r}_{i,u} \cdot  {\bf A}  
  -\delta I^T_i \int \limits_{S_i} du~dv   
~{\bf r}_{i,v} \cdot  {\bf A} 
  - \int \limits_{S_i} du~dv ~\delta \phi_i~( {\bf r}_{i,v} \times {\bf r}_{i,u})
\cdot {\bf B} \Big ) \nonumber  \\ 
 &-&(F^\prime_T~ \delta I^P_1 + F^\prime_P~ \delta I^T_1)\! \int \limits_{S_1} du~dv 
\nabla s \cdot {\bm \xi}\! 
+\! \int \limits_{S_1}\! du ~dv~ \delta \phi_1~ ( F^\prime_T \frac{\partial}{\partial v}
+F^\prime_P \frac{\partial}{\partial u}) \nabla s  \cdot {\bm \xi} \\
 &+&\delta I^P_2  \!\int \limits_{S_2}\! du~dv \frac{{\bf r}_{2,u} \cdot {\bf j}_2}{\gamma \sigma d}  
-\delta I^T_2  \!\int \limits_{S_2}\! du~dv \frac{{\bf r}_{2,v} \cdot {\bf j}_2}{\gamma \sigma d} 
  -\! \int \limits_{S_2}\! du~dv ~\delta \phi_2 \frac{{\bf r}_{2,v} \cdot {\bf j}_{2,u}
-{\bf r}_{2,u} \cdot {\bf j}_{2,v}}{\gamma \sigma d}\nonumber  
\end{eqnarray}
 with
\begin{eqnarray}
 {\bf A}({\bf r})  & =& \!\frac{1}{4\pi} \sum_{k=1}^3 \int \limits_{S_k} \! df_k 
\frac{{\bf j}_k } {\mid {\bf r}-{\bf r}_k \mid}
\end{eqnarray}
From 
 ${\delta \cal L}_V= 0$ one gets
 the boundary conditions at the plasma-vacuum interface $S_1$ 
\begin{eqnarray}   \label{104a}
  \int \limits_{S_1} du~dv   
~{\bf r}_{1,u} \cdot  {\bf A}({\bf r}_1) & =& F^\prime_T~  
  \int \limits_{S_1} du~dv (\nabla s  \cdot {\bm \xi}), \nonumber \\ 
   \int \limits_{S_1} du~dv   
~{\bf r}_{1,v} \cdot  {\bf A}({\bf r}_1) &=& -F^\prime_P~
  \int \limits_{S_1} du~dv (\nabla s  \cdot  {\bm \xi}) \\
  ( {\bf r}_{1,v} \times {\bf r}_{1,u})
\cdot {\bf B}({\bf r}_1) 
 &=& ( F^\prime_T \frac{\partial}{\partial v}
+F^\prime_P \frac{\partial}{\partial u}) (\nabla s  \cdot {\bm \xi}) \nonumber 
\end{eqnarray}
 at the resistive wall $S_2$
\begin{eqnarray}  \label{105a}
  \int \limits_{S_2} du~dv   
~{\bf r}_{2,u} \cdot  {\bf A}({\bf r}_2) & =&   
 \int \limits_{S_2}\! du~dv \frac{{\bf r}_{2,u} \cdot {\bf j}_2}{\gamma \sigma d} \nonumber\\ 
   \int \limits_{S_2} du~dv   
~{\bf r}_{2,v} \cdot  {\bf A}({\bf r}_2) &=& 
  \!\int \limits_{S_2}\! du~dv \frac{{\bf r}_{2,v} \cdot {\bf j}_2}{\gamma \sigma d}  \\
  ( {\bf r}_{2,v} \times {\bf r}_{2,u})
\cdot {\bf B}({\bf r}_2)  &=&
   \frac{{\bf r}_{2,v} \cdot {\bf j}_{2,u}
-{\bf r}_{2,u} \cdot {\bf j}_{2,v}}{\gamma \sigma d}\nonumber  
\end{eqnarray}
and at the external ideal wall $S_3$
\begin{equation}  \label{106a}
  \int \limits_{S_3} du~dv   
~{\bf r}_{3,u} \cdot  {\bf A}({\bf r}_3)  = 0 ,~~
   \int \limits_{S_3} du~dv   
~{\bf r}_{3,v} \cdot  {\bf A}({\bf r}_3) =  0,~~ 
  ( {\bf r}_{3,v} \times {\bf r}_{3,u})
\cdot {\bf B}({\bf r}_3)  = 0
\end{equation} 
 The boundary conditions (A.10,A.11,A.12) obtained for ${\bf A}$ are integrals. With an 
 appropriate  gauge ${\bf A} -> {\bf A}+\nabla \Psi $ one can get the local
conditions (\ref{206}).

 As shown before,  on a toroidal surface there exists for an arbitrary  prescribed net-poloidal(net-toroidal) current $I^P$
($I^T$) a so-called ´'equilibrium'  current distribution ${\bf j}^P_{eq}$ 
( ${\bf j}^T_{eq}$ )
generating a magnetic field with vanishing normal component on the surface.
The field produced by such a poloidal current on $S_1$ and a toroidal current on $S_3$ is zero in the domain bounded by the boundaries $S_1$ and $S_3$.
With  these equilibrium currents one can compensate the $I^P_1$ and $I^T_3$ contributions to the solution.
Therefore one can omit these terms.

%% file: inductance.tex
\par
\medskip
 The  elements of the matrices ${\bf M}_{ij},~i,j=1,2,3$  in (\ref{302})

\begin{equation}  \label{100b}
        {\bf M}_{ij} 
= \left( \begin{array}{cccc}
     g^{ij}_{11} & g^{ij}_{12} & g^{ij}_{1c^\prime} & 
     -g^{ij}_{1s^\prime}  \\ 
     g^{ij}_{21} & g^{ij}_{22} & g^{ij}_{2c^\prime} & 
     -g^{ij}_{2s^\prime} \\ 
      g^{ij}_{c1}  &  
      g^{ij}_{c2} & 
     g^{ij}_{cc^\prime} & 
    -g^{ij}_{cs^\prime}  \\
     -g^{ij}_{s1}  &  
     -g^{ij}_{s2} & 
    -g^{ij}_{sc^\prime} & 
     g^{ij}_{ss^\prime} \\ 
\end{array} \right) 
\end{equation}

 are defined by
\begin{eqnarray}  \label{101b}
  g^{ij}_{cc^\prime}( k, k^\prime)\!\! & =& \!\! \frac{1}{4 \pi}
\int \limits_{S_i} dx \int \limits_{S_j} dx^\prime  
\cos(kx) \frac{(k{\bf R}_{i,x}) \cdot
 (k^\prime~{\bf R}_{j,x^\prime})}{
\mid {\bf r}_i - {\bf r}_j^{\prime} \mid}~ \cos( k^\prime x^\prime)  \nonumber \\
  g^{ij}_{1c^\prime}( k^\prime) & =& -\frac{1}{4 \pi}
\int \limits_{S_i} dx \int \limits_{S_j} dx^\prime  
    \frac{ {\bf r}_{i,v} \cdot
(k^\prime~{\bf R}_{j,x^\prime})}{
\mid {\bf r}_i - {\bf r}_j^{\prime} \mid} \cos( k^\prime x^\prime)   \nonumber \\
  g^{ij}_{2c^\prime}( k^\prime) & =& \frac{1}{4 \pi}
\int \limits_{S_i} dx \int \limits_{S_j} dx^\prime  
    \frac{ {\bf r}_{i,u} \cdot
(k^\prime~{\bf R}_{j,x^\prime})}{
\mid {\bf r}_i - {\bf r}_j^{\prime} \mid} \cos(k^\prime x^\prime)    \\
  g^{ij}_{11}  &=& \frac{1}{4 \pi}
\int \limits_{S_i} dx  \int \limits_{S_j} dx^\prime  
    \frac{ {\bf r}_{i,v} \cdot
{\bf r}_{j,v^\prime}}{
\mid {\bf r}_i - {\bf r}_j^{\prime} \mid}, ~~~ 
  g^{ij}_{12}  = -\frac{1}{4 \pi}
\int \limits_{S_i} dx   \int \limits_{S_j} dx^\prime  
    \frac{ {\bf r}_{i,v} \cdot
{\bf r}_{j,u^\prime}}{
\mid {\bf r}_i - {\bf r}_j^{\prime} \mid}  \nonumber \\
  g^{ij}_{21} & =& -\frac{1}{4 \pi}
\int \limits_{S_i} dx  \int \limits_{S_j} dx^\prime 
    \frac{ {\bf r}_{i,u} \cdot
{\bf r}_{j,v^\prime}}{
\mid {\bf r}_i - {\bf r}_j^{\prime} \mid},~~~ 
  g^{ij}_{22}  =  \frac{1}{4 \pi}
\int \limits_{S_i} dx  \int \limits_{S_j} dx^\prime  
    \frac{ {\bf r}_{i,u} \cdot
{\bf r}_{j,u^\prime}}{
\mid {\bf r}_i - {\bf r}_j^{\prime} \mid}    \nonumber
\end{eqnarray} 
with  the notation
\begin{equation}  \label{102b}
  k = 2 \pi (m,n), ~~  x=(u,v)^\top, dx= du~dv,~~ {\bf R}_{i,x}=(-{\bf r}_{i,v},{\bf r}_{i,u})^\top
~, F=(F^\prime_P,F^\prime_T)^\top
\end {equation}

The matrix elements with sine are defined accordingly.

 The  elements of matrix ${\bf N}_{22}$  in (\ref{302})
\begin{equation}  \label{103b}
        {\bf N}_{22} 
= \left( \begin{array}{cccc}
     h_{11} & h_{12} & h_{1c^\prime} & 
     -h_{1s^\prime}  \\ 
     h_{21} & h_{22} & h_{2c^\prime} & 
     -h_{2s^\prime} \\ 
      h_{c1}  &  
      h_{c2} & 
     h_{cc^\prime} & 
    -h_{cs^\prime}  \\
     -h_{s1}  &  
     -h_{s2} & 
    -h_{sc^\prime} & 
     h_{ss^\prime} \\ 
\end{array} \right) 
\end{equation}
   are defined as
\begin{eqnarray}   \label{104b}
  h_{cc^\prime}(k,k^\prime) & =&  \frac{1}{4 \pi^2}
\int \limits_{S_2} dx   
    \frac{(k {\bf R}_{2,x}) \cdot
(k^\prime {\bf R}_{2,x})}{
\mid {\bf r}_{2,v} \times {\bf r}_{2,u} \mid} \cos ( k x)
 \cos ( k^\prime x) \nonumber
   \\
  h_{1c^\prime}( k^\prime) & =& \frac{1}{2 \pi}
 \int \limits_{S_2} dx 
    \frac{ {\bf r}_{2,v} \cdot
(k^\prime~{\bf R}_{2,x^\prime})}{
\mid {\bf r}_{2,v} \times {\bf r}_{2,u} \mid} 
 \cos (k^\prime x) \nonumber \\
  h_{2c^\prime}( k^\prime) & =& \frac{1}{2 \pi}
 \int \limits_{S_2} dx  
    \frac{ {\bf r}_{2,u} \cdot
(k^\prime~{\bf R}_{2,x^\prime})}{
\mid {\bf r}_{2,v} \times {\bf r}_{2,u} \mid} 
 \cos (k^\prime x)) \\
  h_{11}  &=& 
 \int \limits_{S_2} dx 
    \frac{ {\bf r}_{2,v} \cdot
{\bf r}_{2,v}}{
\mid {\bf r}_{2,v} \times {\bf r}_{2,u} \mid},~~~ 
  h_{12}  = 
 \int \limits_{S_2} dx 
    \frac{ {\bf r}_{2,v} \cdot
{\bf r}_{2,u}}{
\mid {\bf r}_{2,v} \times {\bf r}_{2,u} \mid} \nonumber \\
  h_{21} & =& 
 \int \limits_{S_2} dx 
    \frac{ {\bf r}_{2,u} \cdot
{\bf r}_{2,v}}{
\mid {\bf r}_{2,v} \times {\bf r}_{2,u} \mid},~~~ 
  h_{22}  =  
 \int \limits_{S_2} dx 
    \frac{ {\bf r}_{2,u} \cdot
{\bf r}_{2,u}}{
\mid {\bf r}_{2,v} \times {\bf r}_{2,u} \mid}  \nonumber
\end{eqnarray}
and the matrix elements with sine are defined accordingly.

The  matrix ${\bf M}_{1 \xi}$  (\ref{302})  is defined by 
\footnote{note:  $\delta_{k,k^\prime} = \delta_{m,m^\prime}~\delta_{n,n^\prime}$} 
\begin{equation} \label{105b}
        {\bf M}_{1 \hat \xi} 
= \left( \begin{array}{ccc|c|ccc}
   0 & 0 & 0 & -F^\prime_P & 0 & 0 & 0\\  
   0 & 0 & 0 & -F^\prime_T & 0 & 0 & 0 \\  \hline
   0 & 0 & 0 & 0           &\ddots & 0 & 0\\ 
    0 & 0 & 0 & 0           &0 &- \frac{1}{2} k F \delta_{k,k^\prime} & 0 \\ 
   0 & 0 & 0 & 0           & 0 &   0      & \ddots  \\ \hline
  \ddots & 0 & 0 & 0 &0  & 0 & 0\\ 
   0  & \frac{1}{2} k F \delta_{k,k^\prime}  & 0  & 0 &0  & 0 & 0 \\ 
   0   &    0            & \ddots  & 0 &0 & 0 & 0
\end{array} \right) 
\end{equation}

The matrices ${\bf M}_{  \hat \xi i},i=1,2,3$ in (\ref{309}) are obtained by discretizing the vacuum
energy term

\begin{equation}  \label{106b}
        {\bf M}_{\hat \xi i} 
= \left( \begin{array}{cccc}
     b^i_{s1} & b^i_{s2} & b^i_{ss^\prime} & 
      b^i_{sc^\prime}  \\ 
     \delta_{1,i} F^\prime_P & (\delta_{1,i}-1)F^\prime_T & 0 & 0\\
      b^i_{c1}  &  
      b^i_{c2} & 
     b^i_{cs^\prime} & 
     b^i_{cc^\prime}  \\
\end{array} \right) 
\end{equation}
\begin{eqnarray}  \label{107b}
 b^i_{s1}(k) &=& -\frac{1}{4\pi} 
   \int_{S_i}\!  dx~ \sin(k x)~   d^i_1(u,v), ~~~~
 b^i_{s2}(k) =  \frac{1}{4\pi} 
   \int_{S_i}\!  dx~ \sin(k x)~  d^i_2(u,v)
\end{eqnarray}
  with
\begin{eqnarray}   \label{108b}
  d^1_1(u,v)  &= & 
 (F_P^\prime~ {\bf r}_{1,u}
+ F_T^\prime~ {\bf r}_{1,v}) \cdot
   \dashint_{S_1}\!  du^\prime dv^\prime    
 {\bf r}_{1,v^\prime} \times \frac{({\bf r}_1-{\bf r}_1^\prime)}{|{\bf r}_1-{\bf r}_1^\prime|^3} 
  -2 \pi ~ F^\prime_P~  \nonumber\\
  d^i_1(u,v)  &= & 
 (F_P^\prime~ {\bf r}_{1,u}
+ F_T^\prime~ {\bf r}_{1,v}) \cdot
   \int_{S_i}\!  du^\prime dv^\prime    
 {\bf r}_{i,v^\prime} \times \frac{({\bf r}_1-{\bf r}_i^\prime)}{|{\bf r}_1-{\bf r}_i^\prime|^3}, ~~ i=2,3   \\
  d^1_2(u,v)  &= & 
 (F_P^\prime~ {\bf r}_{1,u}
+ F_T^\prime~ {\bf r}_{1,v}) \cdot
   \dashint_{S_1}\!  du^\prime dv^\prime    
 {\bf r}_{1,u^\prime} \times \frac{({\bf r}_1-{\bf r}_1^\prime)}{|{\bf r}_1-{\bf r}_1^\prime|^3} 
  +2 \pi ~ F^\prime_T~  \nonumber  \\
  d^i_2(u,v)  &= & 
 (F_P^\prime~ {\bf r}_{1,u}
+ F_T^\prime~ {\bf r}_{1,v}) \cdot
   \int_{S_i}\!  du^\prime dv^\prime    
 {\bf r}_{i,u^\prime} \times \frac{({\bf r}_1-{\bf r}_i^\prime)}{|{\bf r}_1-{\bf r}_i^\prime|^3} ,~~ i=2,3  \nonumber
\end{eqnarray}

 and with $F=(F^\prime_P,F^\prime_T)^\top$
\begin{eqnarray}   \label{109b}
 b^i_{ss^\prime}(k,k^\prime) &=&  -\frac{1}{2}(k F) \int_{S_i}\!  dx    
  ~ \cos(k x)~ a^i_{s^\prime }(u,v,k^\prime)  \nonumber\\
 b^i_{cc^\prime}(k,k^\prime) &=&  -\frac{1}{2}(k F) \int_{S_i}\!  dx  
  ~ \sin(k x)~ a^i_{c^\prime }(u,v,k^\prime)  \\
 b^i_{sc^\prime}(k,k^\prime) &=&  -\frac{1}{2}(k F) \left 
( \pi ~\delta_{k,k^\prime}~\delta_{1,i}  + \int_{S_i}\!  dx   
 ~  \cos(k x)~ a^i_{c^\prime }(u,v,k^\prime) \right) \nonumber \\
 b^i_{cs^\prime}(k,k^\prime) &=&   \frac{1}{2}(k F) \left 
( \pi ~\delta_{k,k^\prime}~\delta_{1,i}  + \int_{S_i}\!  dx    
 ~  \sin(k x)~ a^i_{s^\prime }(u,v,k^\prime) \right) \nonumber 
\end{eqnarray}
 with
\begin{eqnarray}   \label{110b}
 a^1_{s^\prime}(u,v,k^\prime) &=&  \dashint_{S_1}\!  dx^\prime     
  ({\bf r}_{1,v^\prime} \times {\bf r}_{1,u^\prime}) \cdot \frac{({\bf r}_1-{\bf r}_1^\prime)}{|{\bf r}_1-{\bf r}_1^\prime|^3}~\sin(k^\prime x^\prime)  \\ 
 a^i_{s^\prime}(u,v,k^\prime) &=&  \int_{S_1}\!  dx^\prime     
  ({\bf r}_{1,v^\prime} \times {\bf r}_{1,u^\prime}) \cdot \frac{({\bf r}_i-{\bf r}_1^\prime)}{|{\bf r}_i-{\bf r}_1^\prime|^3}~\sin(k^\prime x^\prime) ,~i=2,3   \nonumber
\end{eqnarray}
and $a^i_{c^\prime}(u,v,k^\prime)$ accordingly.

%% file: subtract.tex
The elements of the self-inductance matrix ${\bf M}_{11}$  (\ref{100b}) and 
(\ref{101b}) are Fourier integrals over the plasma-vacuum interface.
The integrands  are singular
so that a standard numerical Fourier transform is not possible.
The problem is solved by applying a subtraction method:
An analytically  integrable function with the same singular behaviour
is subtracted. The analytical integral is then added to the numerically
Fourier-transformed regularized  function.

The subtraction method will be   demonstrated by treating a  term of (\ref{101b}).
\begin{equation} 
 \hat g(m,n;m^\prime,n^\prime) = 
  \int \limits_0^1 du \int \limits_0^1  dv 
  \int \limits_0^1 du^\prime \int \limits_0^1 dv^\prime   ~
        g(u,v;u^\prime,v^\prime) 
 ~e^{2 \pi i(mu+nv-m^\prime u^\prime-n^\prime v^\prime)}
\end{equation} 
with
\begin{equation}
  g(u,v;u^\prime,v^\prime) = \frac{{\bf r}_u \cdot {\bf r}_{u^\prime}}
   { \mid {\bf r}(u,v) -{\bf r}(u^\prime,v^\prime) \mid}
\end{equation}
Expanding $g$ at the singularity 
\begin{eqnarray} 
{\bf r}^\prime-{\bf r} &=& {\bf r}_u~ \delta u +{\bf r}_v ~\delta v
 +{\bf r}_{uu}~ \frac{{\delta u}^2}{2} 
  +{\bf r}_{uv}~ \delta u~ \delta v
  +{\bf r}_{vv} ~\frac{{\delta v}^2}{2}  \nonumber 
\end{eqnarray} 
with $\delta u =u^\prime -u,~~
\delta v= v^\prime-v$  and 
replacing $\delta u, \delta v$ by $\tan( \pi~ \delta u) / \pi, 
\tan( \pi~ \delta v)/\pi$ one gets a periodic function which can be 
Fourier-transformed analytically with respect to $u^\prime$ and $v^\prime$.
\begin{eqnarray} 
 && g_{sing}(u,v;u^\prime- u,v^\prime-v)   \\
& =&   \frac{\pi~ {\bf r}^2_u }
  { \sqrt{ {\bf r}^2_u \tan^2(\pi(u^\prime\!\!-\!\!u)) +2~ {\bf r}_u \!\cdot\! {\bf r}_v 
   \tan(\pi(u^\prime\!\!-\!\!u)) \tan(\pi(v^\prime\!\!-\!\!v))
     +  {\bf r}_v^2 \tan^2(\pi(v^\prime\!\!-\!\!v)) } }   \nonumber
\end{eqnarray} 

With the analytically  computable integrals  (see appendix D) 

\begin{equation}
I(m,n;a,b,c) =  \int\limits_{0}^{1} du \int \limits_0^1 dv
  \frac{\pi~ e^{2 \pi i (mu+nv)}}
{\sqrt{ a~\tan^2(\pi  u)  +2~b 
\tan(\pi  u) \tan(\pi v)
+ c~ \tan^2(\pi v)}} 
\end{equation}
one  finds for the singular integral
\begin{equation}
 \hat g_{sing}(m,n;m^\prime,n^\prime) = 
\int \limits_0^1 du  \int \limits_0^1 dv~ 
  {\bf r}_u^2 ~ I(m^\prime, n^\prime;{\bf r}_u^2,{\bf r}_u \cdot {\bf r}_v,{\bf r}_v^2) ~
   e^{ 2 \pi i ((m-m^\prime) u+ (n - n^\prime) v)} 
\end{equation}
  The regularized part can  be Fourier transformed numerically
\begin{eqnarray} 
 & &\hat g_{reg}(m,n;m^\prime,n^\prime) = \\
 &=& \int \limits_0^1 du \int \limits_0^1  dv 
  \int \limits_0^1 du^\prime \int \limits_0^1 dv^\prime   ~
(
g(u,v;u^\prime,v^\prime) 
 -g_{sing}(u,v;u^\prime-u,v^\prime-v))
 ~e^{2 \pi i(mu+nv-m^\prime u^\prime-n^\prime v^\prime)} \nonumber
\end{eqnarray} 
so that the  Fourier transform of $g(u,v;u^\prime,v^\prime)$ is given by
\begin{equation}
 \hat g(m,n;m^\prime,n^\prime) = 
 \hat g_{reg}(m,n;m^\prime,n^\prime) + 
 \hat g_{sing}(m,n;m^\prime,n^\prime)  
\end{equation}

A second type of singular integrals appears in a part of the vacuum energy matrix
${\bf M}_{\hat \xi 1}$ (\ref{106b}) -(\ref{110b}).
The subtraction method  will be 
demonstrated for a term in (\ref{110b}) 
\begin{equation} 
 \hat h(u,v;m^\prime,n^\prime) = 
  \int \limits_0^1  du^\prime \int \limits_0^1 dv^\prime   
        h(u,v;u^\prime,v^\prime) 
 ~e^{2 \pi i(m^\prime u^\prime+n^\prime v^\prime)}
\end{equation} 
with
\begin{equation}
  h(u,v;u^\prime,v^\prime) = \frac{({\bf r}_{v^\prime} \times {\bf r}_{u^\prime}) \cdot (
{\bf r}(u,v)-{\bf r}(u^\prime,v^\prime))}
   { \mid {\bf r}(u,v) -{\bf r}(u^\prime,v^\prime) \mid^3}
\end{equation}
the expansion  of the  integrand at the singularity is given by
\begin{eqnarray}
 &  &  h_{sing}(u,v;u^\prime- u,v^\prime-v) = \\ 
 & &   \frac{\pi ({\bf r}_v \times {\bf r}_u)}{2} \cdot
 \frac{ {\bf r}_{uu} \tan^2(\pi(u^\prime\!\!-\!\!u)) +2~ {\bf r}_{uv} \!  
   \tan(\pi(u^\prime\!\!-\!\!u))\tan(\pi(v^\prime\!\!-\!\!v))
     +  {\bf r}_{vv} \tan^2(\pi(v^\prime\!\!-\!\!v)) } 
  { {( {\bf r}^2_u \tan^2(\pi(u^\prime\!\!-\!\!u))+2~{\bf r}_u \!\cdot\! {\bf r}_v 
   \tan(\pi(u^\prime\!\!-\!\!u)) \tan(\pi(v^\prime\!\!-\!\!v))
     +  {\bf r}_v^2 \tan^2(\pi(v^\prime\!\!-\!\!v)) })^{3/2} } \nonumber
\end{eqnarray}
With the analytically  computable integrals (see apendix D) 
\begin{eqnarray}
 & & K(m,n;A,B,C,a,b,c) =  \\
 &=& \pi \int\limits_{0}^{1} du \int \limits_0^1 dv
\frac{( A\tan^2(\pi  u)  +2B 
\tan(\pi  u) \tan(\pi v)
+ C \tan^2(\pi v)) 
   e^{2 \pi i (m u+n v)}}
{( a~\tan^2(\pi  u)  +2~b 
\tan(\pi  u) \tan(\pi v)
+ c~ \tan^2(\pi v))^{3/2}}  \nonumber
\end{eqnarray}
one gets for the singular integral
\begin{equation}
 \hat h_{sing}(u,v;m^\prime,n^\prime) = 
   K(m^\prime, n^\prime;A,B,C ,{\bf r}_u^2,{\bf r}_u \cdot {\bf r}_v,{\bf r}_v^2) ~
   e^{ 2 \pi i (m^\prime u+ n^\prime v)} 
\end{equation}
with $
 A =  ({\bf r}_v\! \times {\bf r}_u) \cdot {\bf r}_{uu},~ 
 B =  ({\bf r}_v \times {\bf r}_u) \cdot {\bf r}_{uv},~
 C =  ({\bf r}_v \times {\bf r}_u) \cdot {\bf r}_{vv}  $.

The regularized integral can be Fourier transformed numerically.
\begin{equation} 
 \hat h_{reg}(u,v;m^\prime,n^\prime)  
 =   \int \limits_0^1 du^\prime \int \limits_0^1 dv^\prime 
(
h(u,v;u^\prime,v^\prime) 
 -h_{sing}(u,v;u^\prime-u,v^\prime-v))
 ~e^{2 \pi i(m^\prime u^\prime+n^\prime v^\prime)} 
\end{equation} 
For the Fourier transform of $h(u,v;u^\prime,v^\prime)$ one gets
\begin{equation}
 \hat h(u,v;m^\prime,n^\prime) = 
 \hat h_{reg}(u,v;m^\prime,n^\prime) + 
 \hat h_{sing}(u,v;m^\prime,n^\prime)  
\end{equation}

%% file: regul_integrals.tex
\par
\medskip
In appendix C a subtraction method to regularize singular Fourier integrals
has been presented using the following analytically computable integrals
 \cite{Merkel1986},\cite{Merkel1987} :
\begin{equation}  \label{100d}
I_{mn}=\pi\int\limits_0^1\int\limits_0^1\,du\,dv\,
{
e^{2\pi i(mu+nv)}\over
(a\tan^2(\pi u)+2b\tan(\pi u)\tan(\pi v)+
c\tan^2(\pi v))^{1\over2}}
\end{equation}
and 
\begin{equation} \label{101d}
K_{mn}=\pi\int\limits_0^1\int\limits_0^1\,du\,dv\,
{(A\tan^2(\pi u)+2B\tan(\pi u)\tan(\pi v)+C\tan^2(\pi v))
e^{2\pi i(mu+nv)}\over
(a\tan^2(\pi u)+2b\tan(\pi u)\tan(\pi v)+
c\tan^2(\pi v))^{3\over2}}
\end{equation}
with $ac - b^2 > 0 $.
The integrals $K_{mn}$ can be obtained by deriving  the integrals
$I_{mn}$ with respect to $a,b,c$ :
\begin{equation} \label{102d}
K_{mn}=-2(A{\partial\over \partial a}+B{\partial\over
\partial b}+C{\partial\over \partial c})I_{mn}\,.
\end{equation}
To compute the $I_{mn}$ a generating function ${\cal I}$ is
introduced:
\begin{equation} \label{103d}
{\cal I}=\sum\limits_{m=0,n=0}^{\infty ,\infty}
I_{mn}s^mt^n\,.
\end{equation}
Summing up the power series, one obtains ${\cal I}$ in
closed form:
\begin{equation} \label{104d}
{\cal I}=\pi\int\limits_0^1\int\limits_0^1
{du\,dv \over (1-se^{2\pi iu})(1-te^{2\pi iv})
(a\tan^2(\pi u)+2b\tan(\pi u)
\tan(\pi v)+c\tan^2(\pi v))^{1\over2}}.
\end{equation}
With the variables $(r, y): ~ (
y = \tan \pi u,~~
ry= \tan \pi v)
$
one gets for ${\cal I}$
\begin{equation}   \label{105d}
{\cal  I} = {1 \over 4\pi} \int\limits_{-\infty}^{+\infty}
\int\limits_{-\infty}^{+\infty}
\,dr\,dy\,\biggl( {1 \over y+i\alpha}-{1 \over y-i}
\biggr) \biggl( {1 \over ry+i\beta} - {1 \over ry-i}
\biggr) {1 \over (a+2br+cr^2)^{1 \over 2}}
\end{equation}
with
\begin{equation} \label{106d}
\alpha  = {1-s \over 1+s},~~ 
\beta  = {1-t \over 1+t} .
\end{equation}
Integrating ${\cal I}$ with respect to $y$, one obtains
\begin{eqnarray} \label{107d}
{\cal I} &=& {1 \over 2} \int\limits_0^\infty\,dr
\biggl( {1 \over \beta +\alpha r} + {1 \over 1+r}
\biggr) {1 \over (a-2br+cr^2)^{1 \over 2}} \\
&+& {1 \over 2} \int\limits_0^\infty\,dr
\biggl( {1 \over 1+\alpha r} + {1 \over \beta +r}
\biggr) {1 \over (a+2br+cr^2)^{1 \over 2}}\nonumber \,.
\end{eqnarray}
and
\begin{eqnarray} \label{108d}
   \int\limits_0^\infty\,dr
\biggl( {1 \over \beta +\alpha r}
\biggr) {1 \over \sqrt{a-2br+cr^2}} &=&
   \frac{1}{\gamma} \ln{\frac{\alpha(\gamma \sqrt{c} +\alpha c + 
  \beta b)}{\beta (\gamma \sqrt{a} -\alpha b - \beta a)}} 
\end{eqnarray}
with $\gamma = \sqrt{a \beta^2 + 2 b \alpha \beta +c \alpha^2}$

With the substitution $x:~ x = (1-r)/(1+r)$ 
the function ${\cal I}$ can be written as a sum of four terms:
\begin{equation} \label{109d}
{\cal I} = h^+(s,t)+h^+(0,0)+h^-(s,0)+h^-(0,t)\,,
\end{equation}
with
\begin{equation} \label{110d}
h^\pm (s,t)={(1+s)(1+t) \over 2}\,
\int\limits_{-1}^{+1}\,{dx \over (1-st-(s-t)x)
(a^\mp +2dx+a^\pm x^2)^{1 \over 2}}
\end{equation}
and 
\begin{eqnarray*}
a^+ &=& a+2b+c, \\
a^- &=& a-2b+c, \\
d &=& c-a  .
\end{eqnarray*}
The $I_{mn}$ are then obtained by expanding
${\cal I}$ again as a power series in $s$ and $t$. One starts
by expanding the functions $h^\pm$
\begin{equation} \label{111d}
h^\pm (s,t)={1 \over 2}
{(1+s)(1+t) \over 1-st} \sum\limits_{\ell =0}^\infty
\biggl( {s-t \over 1-st}\biggr)^\ell T_\ell ^\pm
\end{equation}
with
\begin{equation} \label{112d}
T_\ell ^\pm = \int\limits_{-1}^{+1}\,dx\,
{x^\ell \over (a^\mp+2dx+a^\pm x^2)^{1 \over 2}}
\end{equation}
Expanding the  $h^\pm$ further, one finally obtains 
\begin{eqnarray} \label{113d}
I_{mn}=\left\{\begin{array}{*{3}{c@{+}}c@{,}c} 
 c^+_{m\,n} &c^+_{m-1\,n} &c^+_{m\,n-1} &c^+_{m-1\,n-1}&
~{\rm for}\, m\ge 1,n\ge 1\\
c^+_{m\,0} &c^+_{m-1\,0} &c^-_{m\,0} &c^-_{m-1\,0}&
~{\rm for}\, m\ge 1,n=0\\
c^+_{0\,n} &c^+_{0\,n-1} &c^-_{0\,n} &c^-_{0\,n-1}&
~{\rm for}\, m=0,n\ge 1\\
c^+_{0\,0} &c^+_{0\,0} &c^-_{0\,0} &c^-_{0\,0}&
~{\rm for}\, m=0,n=0\end{array} \right.
\end{eqnarray}
with
$$
c^\pm_{mn}=\sum\limits_{\ell =0}^{{m+n-\vert m-n\vert \over 2}}
{(-1)^{\ell +{\vert m-n\vert -m+n \over 2}}
\bigl({m+n+\vert m-n\vert
\over 2}+\ell \bigr) !\,  T_{\vert m-n\vert +2\ell}^\pm
\over 2 \bigl({m+n-\vert m-n\vert \over 2}-\ell \bigr) !
(\vert m-n\vert +\ell)! \ell !}\,.
$$
The integrals $T_\ell ^\pm$  can be
calculated by using a recurrence relation

\begin{eqnarray} \label{114d}
T_0^\pm & = &{1 \over \sqrt{a^\pm}}\log
{\sqrt {c~ a^\pm }+c\pm b \over
\sqrt {a~ a^\pm}-a\mp b}, \nonumber\\
T_1^\pm &= &{1 \over {a^\pm }}
\bigl( 2(\sqrt c -\sqrt a)-(c-a)T_0^\pm \bigr),  \\
T_\ell ^\pm &= &{1 \over {\ell ~ a^\pm }}
\bigl( 2(\sqrt c +(-1)^\ell \sqrt a)   
  -   (2\ell -1)(c-a)T_{\ell -1}^\pm -(\ell -1)~
a^\mp~ T_{\ell -2}^\pm\bigr)\quad
{\rm for}\  \ell \ge 2\,. \nonumber
\end{eqnarray}
The Fourier integrals $K_{mn}$ follow from the $I_{mn}$
by differentiation. 
One gets the same formulas (D.14) as for
the $I_{mn}$ replacing the integrals $T_\ell ^\pm$
 by the integrals $S_\ell ^\pm$, which are
given by
\begin{equation} \label{115d}
S_\ell ^\pm = \int\limits_{-1}^{+1}\,dx\,x^\ell\,
{(A^\mp+2D
\,x+A^\pm x^2) \over (a^\mp+2d\,x+a^\pm x^2)^{3 \over 2}}
\end{equation}
with
\begin{eqnarray*}
A^+ &=& A+2B+C, \\
A^- &=& A-2B+C, \\
D &=& C-A \ .
\end{eqnarray*}
As for the $T_\ell ^\pm$, recurrence formulas can be derived
for the $S_\ell ^\pm$, but the $S_\ell ^\pm$ can also be
expressed in terms of the $T_\ell ^\pm$ by appropriate partial
integration. One obtains for $S_\ell ^+$
\begin{eqnarray} \label{116d}
 &  & (a+2b+c)   (ac-b^2)S_\ell ^+ =  \nonumber\\
& &\biggl( (A+2B+C)(ac-b^2)+\ell \bigl( k_1(a+2b+c)+k_2(c-a)
\bigr) \biggr)\,T_\ell ^+ \nonumber \\
&+&\ell \bigl(k_1(c-a)+k_2(a-sb+c) \bigr)\,T_{\ell -1}^+ \\
&-& {(c+b)k_1+(c-b)k_2 \over \sqrt c}\,-\,(-1)^\ell\,
{(a+b)k_1-(a-b)k_2 \over \sqrt a}   \nonumber
\end{eqnarray}
with
\begin{eqnarray*}
k_1 &=& (a+c)B-(A+C)b, \\
k_2 &=& C(a+b)-B(c-a)-A(c+b). 
\end{eqnarray*}
The formula for $S^- _\ell$ is obtained by replacing
$b$ by   $-b$ and $B$ by   $-B$, and $T^+ _\ell , T^+ _{\ell -1}$
 by  $T^- _\ell , T^- _{\ell -1}$.
\par
\hskip.5truecm

The forward computation of the recurrence relation $T^+_{\ell}$ gets unstable 
if $b<0$,  and for $T^-_{\ell}$ if $b>0$ \cite{Wimp1984}.
Considering the homogeneous equation
\begin{equation} \label{117d}
  {\cal L}[T^+_\ell] \equiv 
T^+_{\ell+2} +\frac{(2\ell+3) d}{(\ell+2) a^+} ~T^+_{\ell+1} +\frac{(\ell+1) a^-}{
 (\ell+2) a^+}~ T^+_\ell = 0
\end{equation}
with the ansatz $ T^+_{\ell} \approx t^{\ell}$ one gets the characteristic
equation 
\begin{equation} \label{118d}
t^2 +\frac{(2\ell+3) d}{(\ell+2) a^+} ~t +\frac{(\ell+1) a^-}{
 (\ell+2) a^+}~  = 0
\end{equation}
with the solution for large $\ell \rightarrow \infty$
\begin{equation} \label{119d}
  t_\pm = -\frac{d}{a^+} \pm i \frac{\sqrt{a~c-b^2}}{a^+} ~ \mbox{and} ~\mid t_\pm \mid=
 \frac{a^-}{a^+} 
\end{equation}
For $b<0$ the moduli are $\mid t_\pm \mid > 1$ so that the forward recursion
is unstable.
In that case the $T^+_\ell$ can be obtained by solving the recurrence relation
in the backward direction.
First one generates a solution of the nonhomogeneous equation by using the equation in the backward direction with starting values
$$
z_N(N+1) = z_N(N)= 0.
$$
Then one generates a solution $y_N(\ell)$ of the homogeneous equation
${\cal L}[y(\ell)]=0$ by using the equation in the backward direction
with
$$
 y_N(N+1) = 0, ~~y_N(N)=1.
$$
Then the solution $w_N(\ell)$ is obtained by
\begin{equation} \label{120d}
w_N(\ell) = \lambda(N)~ y_N(\ell) + z_N(\ell),~~~ 0 \le \ell \le N+1 
\end{equation}
with
$$
 \lambda(N) = \frac{T^+_0-z_N(0)}{y_N(0)}
$$
For $N \rightarrow \infty$  the $w_N(\ell)$ converges to $T^+_{\ell}$.

%% file: asymptotic.tex
\par 
\medskip
For large values of  $m,n$ an asymptotic expansion for the Fourier
integral (\ref{100d}) 
\begin{eqnarray} 
I_{mn}=\pi \int\limits_{-\frac{1}{2}}^{\frac{1}{2}}
          \int\limits_{-\frac{1}{2}}^{\frac{1}{2}}\,du\,dv\,
\frac{e^{2 \pi i(mu+nv)}}{
(a \tan^2(\pi u)+2b \tan(\pi u)\tan(\pi v)+
c \tan^2(\pi v))^{\frac{1}{2}}}\,,
\end{eqnarray}
can be derived.

With the substitution 
\[
    \pi u  =  r~ \cos \varphi,~~ 
    \pi v  =  r~ \sin \varphi  
\]
and
\begin{equation} \label{121d}
             m  =  \lambda \cos \varphi_0,~~  
             n  =  \lambda \sin \varphi_0 , ~~ \lambda^2  = m^2+n^2 
\end{equation}
one gets
\begin{eqnarray}  \label{122d}
I_{mn}= \int\limits_{0}^{2\pi} d\varphi \int\limits_{0}^{R(\varphi)}\,dr~
 g(r,\varphi)~ e^{  i \lambda f(r,\varphi)} 
\end{eqnarray}
 with
\begin{eqnarray*}
 f(r,\varphi)~ &=& 2 r \cos(\varphi- \varphi_0) \\
g(r,\varphi) &=&
     \frac{r}{\pi (a~ \tan^2(r \cos \varphi)+2b \tan(r \cos \varphi)
 \tan(r \sin \varphi) +c \tan^2(r \sin \varphi)^{1/2})} \nonumber
\end{eqnarray*}
and 
\[
 R(\varphi)= \left \{ \begin{array}{c} 
                    \frac{1}{2 \mid \cos(\varphi) \mid} \\ 
                    \frac{1}{2 \mid \sin(\varphi) \mid}  
              \end{array} 
        \mbox{for}  \mid \varphi-l \pi/2 \mid < \pi/4 
              \begin{array}{c} l=0,2 \\ l=1,3 
              \end{array}
        \right \}
\]
The procedure used is based on the method given in \cite{Wong1989}.
In the present case the main contribution to the asymptotic expansion
comes from the stationary point, where the derivatives
$ \frac{\partial}{\partial r} f(r,\varphi)=0,~ \frac{\partial}{\partial \varphi} f(r,\varphi) =0 $
vanish. 
The stationary point is found to be 
\begin{eqnarray} \label{123d}
      r&=& 0 , \nonumber \\
  \cos( \varphi -\varphi_0)&=& 0 ~~\rightarrow \varphi -\varphi_0 =  \frac{\pi}{2}.
\end{eqnarray}
Defining $\varphi^\prime$ : 
$  \varphi^\prime= \varphi -\varphi_0 -\frac{\pi}{2} $ one gets with
$ \hat g(r,\varphi^\prime) = g(r,\varphi) $ and
$ \hat f(r,\varphi^\prime) = f(r,\varphi) $ 

\begin{eqnarray} \label{124d}
I_{mn}= \int\limits_{-R}^{R} dr \int\limits_{-\frac{\pi}{2}}^{\frac{\pi}{2}}\,d\varphi^\prime ~
 \hat g(r,\varphi^\prime)~ e^{2 i \lambda r \sin(\varphi^\prime)} 
\end{eqnarray}
 with
\[
\hat g(r,\varphi^\prime)\! =\! 
     \frac{r}{\pi (a \tan^2(r \sin(\varphi^\prime\!+\!\varphi_0))
-2b \tan(r \sin(\varphi^\prime\!+\!\varphi_0)) 
 \tan(r \cos(\varphi^\prime\!+\!\varphi_0)) +c \tan^2(r \cos( \varphi^\prime\!+\! \varphi_0))^{1/2}}
\]
There  has been  made  use  of the fact that $\hat g(-r,\varphi)=-\hat g(r,\varphi)$ and
 $\hat g(r,\varphi+\pi)=\hat g(r,\varphi)$ .
For $R$ it is sufficient to choose $R>0$ small but finite.
\vskip.5truecm
Substituting  $\varphi^\prime$ by $t=\sin(\varphi)$ one gets
\begin{eqnarray} \label{125d}
I_{mn}= \int\limits_{-R}^{R} dr \int\limits_{-1}^{1}\,dt~
 G(r,t)~ e^{2 i \lambda r t } 
\end{eqnarray}
 with
\[
G(r,t)=  
     \frac{r}{\pi (1-t^2)^{1/2}(a~ \tan^2(r T_s)
-2b~\tan(r T_s ) 
 \tan(r T_c) +c~\tan^2(r T_c))^{1/2}}
\]
and
\begin{eqnarray}  \label{126d}
  T_s&=& ~~t~ \cos\varphi_0 + \sqrt{1-t^2}~ \sin \varphi_0 , \\
  T_c&=& -t~ \sin \varphi_0 + \sqrt{1-t^2}~ \cos \varphi_0 .  \nonumber
\end{eqnarray}
Substituting 
$   r  =  x-y,~~     
   t    =  x+y
$
one gets
\begin{eqnarray} \label{127d}
I_{mn}=2 \int\limits_{}^{}  dx \int\limits_{}^{}\,dy~
 G(x-y,x+y))~ e^{2 i \lambda (x^2-y^2) }. 
\end{eqnarray}
The asymptotic expansion obtained for this case in 
\cite{Wong1989} is given by
\begin{eqnarray} \label{128d}
I_{mn } &=& -2 \pi ~\sum_{\nu=0}^{\nu_{max}} c_\nu~  e^{i\frac{\pi}{2}\nu} \frac{\nu !}{(2 \lambda)^{\nu+1}},~~
 c_\nu = \sum_{j=0}^\nu F_{2j,2(\nu-j)} ~C_{j, \nu-j} \\
 C_{j,\ell}&=& \frac{(-1)^{\ell+1}}{2^{2j+1}}
\left ( 
  \begin{array}{c}
     2j     \\
     j 
 \end{array}
\right )
\frac{1 \cdot 3 \cdot \cdot \cdot (2\ell-1)}{(2\ell+2j)(2\ell+2j-2)\cdot \cdot \cdot  (2j+2)}    \nonumber \\
  F_{i, k} &=& \frac{1}{i!} \frac{1}{k!} \left  [ 
  \frac{\partial^i}{\partial x^i}   
   \frac{\partial^k}{\partial x^k}  G(x-y,x+y) 
    \right  ]_{ x=0 \atop y=0} \nonumber
\end{eqnarray}

 The coefficients $C_{j,\nu-j}$ can be written also  in closed form
\begin{eqnarray} \label{129d}
   C_{j,\nu-j}&=& \frac{(-1)^{\nu-j+1}}{2^{2j+1}} \frac{(2j)!}{j!~j!}
    \frac{Z}{N} ,~~Z= \frac{(2\nu-2j)!}{2^{\nu-j} (\nu-1)!},~~
    N= \frac{2^{\nu-j} \nu!}{j!}
\end{eqnarray}
 so that one finds
\begin{eqnarray} \label{130}
  I_{mn} &= & 
   \sum \limits_{\nu=0}^{\nu_{max}}
   \frac{4 \pi~e^{i \nu \pi/2} }{\nu!~(2 \lambda)^{\nu+1}}
   \sum \limits_{j=0}^\nu 
      \frac{(-1)^{\nu-j}}{2^{2\nu+1}}
        \frac{\nu!}{j!~(\nu-j)!} 
             \frac{\partial^{2j}}{\partial x^{2j}}
             \frac{\partial^{2(\nu-j)}}{\partial y^{2(\nu-j)}}
                G(x-y,x+y)
\end{eqnarray}
or
\begin{eqnarray*}
  I_{mn}&= & 
   \sum \limits_{\nu=0}^{\nu_{max}}
   \frac{4 \pi }{\nu!~(2 \lambda)^{\nu+1}}
               e^{i \nu \pi/2}  
      \frac{1}{2^{2\nu+1}}
           \left ( \frac{\partial^{2}}{\partial x^{2}}-
             \frac{\partial^{2}}{\partial y^{2}}\right )^\nu
                G(x-y,x+y)
\end{eqnarray*}
or
\begin{eqnarray*}
  I_{mn}&= &  
   \sum \limits_{\nu=0}^{\nu_{max}}
   \frac{4 \pi }{\nu!~(2 \lambda)^{\nu+1}}
               e^{i \nu \pi/2}  
      \frac{1}{2^{2\nu+1}} \left [
           \left ( \left(\frac{\partial}{\partial r}
           +\frac{\partial}{\partial t}\right)^2 - 
           \left (-\frac{\partial}{\partial r}
           +\frac{\partial}{\partial t}\right )^2  
             \right )^\nu
                G(r,t)
               \right ]_{ r=0 \atop t=0}
\end{eqnarray*}
 and finally 
\begin{eqnarray} \label{131d}
  I_{mn}&= & 
   \sum \limits_{\nu=0}^{\nu_{max}}
    \frac{2 \pi~ e^{i\nu \pi/2}}  
          {\nu!~(2 \lambda)^{\nu+1}}
    \left [
    \left ( \frac{\partial}{\partial r}
             \frac{\partial}{\partial t} 
    \right )^\nu G(r,t)
    \right ]_{ r=0 \atop t=0}   .
\end{eqnarray}
The two leading terms of the asymptotic expansion are given by
\begin{eqnarray} \label{132d}
     I_{mn} = \pi  \left [\frac{ G(r,t)}{\lambda}- \frac{G(r,t)_{rrtt}}{8 \lambda^3}
    \right ]_{ r=0 \atop t=0}
\end{eqnarray}
Here it has been made use of the fact that 
 $G(r,t)$ is a function  of $r^2$, so that uneven  derivatives vanish for
$r=0$. 

Expanding $G(r,t)$ at $r=0,t=0$ up to second order in $r$ and $t$  one gets
\begin{eqnarray} \label{133d}
G(r,t)  &=& 
     \frac{1}{ \pi \sqrt{(1-t^2)(  \alpha(t) + \beta(t)~r^2 )}} , \\
 \alpha(t) &=& a~T_s^2-2 b~T_s T_c +c~T_c^2, \nonumber\\
 \beta(t) &=& \frac{2}{3} ( a~T_s^4- b~(T_s T_c^3+ T_c T_s^3) +c~T_c^4). 
 \nonumber
\end{eqnarray}
Using (\ref{126d}) one gets for $\alpha(t)$ and $\beta(t)$ up to second order in
 $t$
\begin{eqnarray} \label{134d}
      \alpha(t) &=&  a_0+a_1~t+a_2~t^2,~~\beta(t)= b_0+b_1~t+b_2~t^2,
\end{eqnarray}
 where the coefficients $a_i,b_i,i=0,1,2$  depend only on the $m,n$-harmonics
 (\ref{121d})
\begin{equation*} 
             \cos \varphi_0 = \frac{m}{\lambda},~~
             \sin \varphi_0 = \frac{n}{\lambda},~~ \lambda = \sqrt{m^2+n^2} ,
\end{equation*}
and are given by
\begin{eqnarray} \label{136d}
      a_0 &=&  a~ \sin^2(\varphi_0) -2 b~ \sin(\varphi_0)~\cos(\varphi_0)
                +c~ \cos^2(\varphi_0), \nonumber\\
  a_1 &=& (a-c)~\sin(2\varphi_0)-2 b~ \cos(2 \varphi_0), \nonumber\\
    a_2   &=&  (a-c)~ \cos(2\varphi_0) +2 b~ \sin(2 \varphi_0), \nonumber\\ 
      b_0 &=& \frac{2}{3}( a~ \sin^4(\varphi_0) - b~ \sin(\varphi_0)~\cos(\varphi_0)
                +c~ \cos^4(\varphi_0)), \\
     b_1  &=&\frac{2}{3} (2~( a~\sin^2(\varphi_0)
                 - c \cos^2(\varphi_0))\sin(2\varphi_0) 
                 - b~ \cos(2 \varphi_0)), \nonumber \\
    b_2   &=&(a+c)~\sin^2(2\varphi_0)-\frac{4}{3}(a~\sin^4(\varphi_0) 
              +c~\cos^4(\varphi_0)) 
                       +\frac{2}{3} b~ \sin(2 \varphi_0). \nonumber
\end{eqnarray}
For the two leading terms of the asymptotic expansion one obtains
\begin{equation} \label{137d}
I_{mn } = \frac{1}{\lambda} \frac{1}{\sqrt{a_0}}
          + \frac{1}{8 \lambda^3}\left ( \frac{b_0+2~b_2}{\sqrt{a_0}^3}
          - 3 \frac{a_2 b_0+a_1 b_1}{\sqrt{a_0}^5}
          + \frac{15}{4} \frac{a_1^2 b_0}{\sqrt{a_0}^7} \right )
\end{equation}
 The asymptotic expansion of  $K_{mn}$   (\ref{101d}) can be obtained from $I_{mn}$  using (\ref{102d})
\begin{eqnarray} \label{138d}
K_{mn } &= &\frac{1}{\lambda} \frac{A_0}{\sqrt{a_0}^3}
          + \frac{3 A_0}{8 \lambda^3}\left ( \frac{b_0+2~b_2}{\sqrt{a_0}^5}
          - 5 \frac{a_2 b_0+a_1 b_1}{\sqrt{a_0}^7}
          + \frac{35}{4} \frac{a_1^2 b_0}{\sqrt{a_0}^9} \right )  \\
         & -& \frac{1}{4 \lambda^3}\left ( \frac{B_0+2~B_2}{\sqrt{a_0}^3}
          - 3 \frac{A_2 b_0+a_2 B_0+A_1 b_1+a_1 B_1}{\sqrt{a_0}^5}
      + \frac{15}{4} \frac{2 a_1 A_1 b_0+a_1^2 B_0}{\sqrt{a_0}^7} \right )
         \nonumber
\end{eqnarray}
where the expressions for $A_i,B_i,i=0,1,2$ are obtained from
(\ref{136d})  by replacing $a,b,c$ by  $A,B,C$.

%% file: paper.bbl
\begin{thebibliography}{10}
\newcommand{\enquote}[1]{`#1'}
\expandafter\ifx\csname url\endcsname\relax
  \def\url#1{{\tt #1}}\fi
\expandafter\ifx\csname urlprefix\endcsname\relax\def\urlprefix{URL }\fi
\providecommand{\eprint}[2][]{\url{#2}}

\bibitem{Chu2010}
Chu M~S and Okabayashi M 2010 \enquote{Stabilization of the external kink and
  resistive wall mode.}
\newblock {\em Plasma Phys. Control. Fusion\/} {\bf 52} 123001.
\newblock \eprint{doi:10.1088/0741-3335/52/12/123001}.

\bibitem{Bialek2001}
Bialek J, Boozer A~H, Mauel M~E, and Navratil G~A 2001 \enquote{Modeling of
  active control of external magnetohydrodynamic instabilities.}
\newblock {\em Phys. Plasmas\/} {\bf 8}(5) 2170--2180.

\bibitem{Bialek2007}
Katsuro-Hopkins O, Bialek J, Maurer D~A, and Navratil G~A 2007
  \enquote{{Enhanced ITER resistive wall mode feedback performance using
  optimal control techniques }.}
\newblock {\em Nucl. Fusion\/} {\bf 47} 1157--1165.

\bibitem{Chu2003}
Chu M, Chance M, Glasser A, and Okabayashi M 2003 \enquote{Normal mode approach
  to modelling of feedback stabilization of the resistive wall mode.}
\newblock {\em Nucl. Fusion\/} {\bf 43} 441--454.

\bibitem{Liu2004}
Liu Y, Bondeson A, Gregoratto D, {\em et~al.\/} 2004 \enquote{Feedback control
  of resistive wall modes in toroidal devices.}
\newblock {\em Nucl. Fusion\/} {\bf 44} 77--86.

\bibitem{Portone2008}
Portone A, Villone F, Liu Y, Albanese R, and Rubinacci G 2008 \enquote{{
  Linearly perturbed MHD equilibria and 3D eddy current coupling via control
  surface method }.}
\newblock {\em Plasma Phys. Control. Fusion\/} {\bf 50} 085004.
\newblock \eprint{doi:10.1088/0741-3335/50/8/085004}.

\bibitem{Hirshman1983}
Hirshman S~P and Whitson J~C 1983 \enquote{Steepest-descent moment method for
  three-dimensional magnetohydrodynamic equilibria.}
\newblock {\em Phys. Fluids\/} {\bf 26} 3553.

\bibitem{Hirshman1986}
Hirshman S~P and Lee D~K 1986 \enquote{Momcon: A spectral code for obtaining
  three-dimensional magnetohydrodynamic equilibria.}
\newblock {\em Comput. Phys. Commun.\/} {\bf 39} 161.

\bibitem{Hirshman1_1986}
Hirshman S~P, van Rij W~I, and Merkel P 1986 \enquote{{Three-dimensional free
  boundary calculations using a spectral Green's function method}.}
\newblock {\em Comput. Phys. Comm.\/} {\bf 43} 143.

\bibitem{Nuehrenberg1996}
N\"uhrenberg C 1996 \enquote{{Global ideal magnetohydrodynamic stability
  analysis for the configurational spcae of Wendelstein 7-X}.}
\newblock {\em Phys. Plasmas\/} {\bf 3} 2401.

\bibitem{Sempf2009}
Sempf M, Merkel P, Strumberger E, Tichmann C, and G{\"u}nter S 2009
  \enquote{Robust control of resistive wall modes using pseudospectra.}
\newblock {\em New J. Phys.\/} {\bf 11} 053015.
\newblock \eprint{doi:10.1088/1367-2630/11/5/053015}.

\bibitem{Hoelzl2012}
H{\"o}lzl M, Merkel P, Huysmans G, {\em et~al.\/} 2012 \enquote{Coupling
  {JOREK} and {STARWALL} codes for non-linear resistive-wall simulations.}
\newblock {\em Journal of Physics: Conference Series\/} {\bf 401} 012010.
\newblock \eprint{doi:10.1088/1742-6596/401/1/012010}.

\bibitem{Huysmans2007}
Huysmans G~T~A and Czarny O 2007 \enquote{{MHD} stability in {X}-point
  geometry: simulation of {ELMs}.}
\newblock {\em Nucl. Fusion\/} {\bf 47} 659.
\newblock \eprint{doi:10.1088/0029-5515/47/7/016}.

\bibitem{Strumberger2014a}
Strumberger E, G{\"u}nter S, Merkel P, and Tichmann C 2014 \enquote{Linear
  stability studies including resistive wall effects with the {CASTOR/STARWALL}
  code.}
\newblock {\em Journal of Physics: Conference Series\/} {\bf 561} 012016.
\newblock \eprint{doi:10.1088/1742-6596/561/1/012016}.

\bibitem{Merkel2004}
Merkel P, N{\"u}hrenberg C, and Strumberger E 2004 \enquote{Resistive wall
  modes of {3D} equilibria with multiply-connected walls.}
\newblock In {\em 31st EPS Conf. on Contr. Fusion and Plasma Phys.\/}. ECA Vol.
  28G P-1.208, London, UK.

\bibitem{Merkel2006}
Merkel P and Sempf M 2006 \enquote{Feedback stabilization of resistive wall
  modes in the presence of multiply-connected wall structures.}
\newblock In {\em Fusion Energy 2006\/}. (Proc. 21st Int. Conf., Chengdu, 2006)
  (Vienna: IAEA), CD-ROM file TH/P3-8, and
  http://www-naweb.iaea.org/napc/physics/FEC/FEC2006/html/index.htm.

\bibitem{Guenter2008}
G{\"u}nter S, Lauber P, Merkel M, {\em et~al.\/} 2008
  \enquote{Three-dimensional effects in tokamaks.}
\newblock {\em Plasma Phys. Control. Fusion\/} {\bf 50} 124004.
\newblock \eprint{doi:10.1088/0741-3335/50/12/124004}.

\bibitem{Strumberger2008}
Strumberger E, Merkel P, Sempf M, and G{\"u}nter S 2008 \enquote{On fully
  three-dimensional resistive wall mode and feedback stabilization
  computations.}
\newblock {\em Phys. Plasmas\/} {\bf 15} 056110.
\newblock \eprint{doi:10.1063/1.2884579}.

\bibitem{Kallenbach2011a}
Kallenbach A, Bobkov V, Braun F, {\em et~al.\/} 2011 \enquote{{ASDEX} {U}pgrade
  results and future plans.}
\newblock In {\em 38th {IEEE} International Conference on Plasma Science
  ({ICOPS}) and 24th Symposium on Fusion Engineering ({SOFE})\/}. SPL5-1,
  Chicago, IL.

\bibitem{Martensen1960}
L\"ust R and Martensen E 1960 \enquote{{Zur Mehrwertigkeit des skalaren
  magnetischen Potentials beim hydromagnetischen Stabilit\"atsproblem eines
  Plasmas}.}
\newblock {\em Z. Naturforschung\/} {\bf 15a} 706.

\bibitem{Merkel1986}
Merkel P 1986 \enquote{{An integral equation technique for the exterior and
  interior Neumann problem in toroidal regions}.}
\newblock {\em J. Comput. Physics\/} {\bf 66} 83.

\bibitem{Merkel1987}
Merkel P 1988 \enquote{{Applications of the Neumann problem to stellarators:
  magnetic surfaces, coils, free-boundary equilibrium, magnetic diagnostic}.}
\newblock {\em Theory of Fusion Plasmas Varenna 1987,EUR 11336 EN\/}  25.

\bibitem{Freidberg1989}
Haney S~W and Freidberg J~P 1989 \enquote{Variational methods for studying
  tokamak stability in the presence of a thin wall.}
\newblock {\em Phys. Fluids B\/} {\bf 1} 1637.

\bibitem{Bernstein1958}
Bernstein I~B, Frieman E~A, Kruskal M~D, and Kulsrud R~M 1958 \enquote{An
  energy principle for hydromagnetic stability problems.}
\newblock {\em Proc. R. Soc. Lond.\/} {\bf 244} 17--40.

\bibitem{Nuehrenberg1987}
N\"uhrenberg J and Zille R 1988 \enquote{Equilibrium and stability of low-shear
  stellarators.}
\newblock {\em Theory of Fusion Plasmas Varenna 1987,EUR 11336 EN\/}  3.

\bibitem{Suttrop2009}
Suttrop W, Gruber O, G{\"u}nter S, {\em et~al.\/} 2009 \enquote{In-vessel
  saddle coils for {MHD} control in {ASDEX} {U}pgrade.}
\newblock {\em Fusion Engineering and Design\/} {\bf 84} 290.
\newblock \eprint{doi:10.1016/j.fusengdes.2008.12.044}.

\bibitem{McAdams2014}
McAdams R 2014 {\em Non-linear Magnetohydrodynamic Instabilities in Advanced
  Tokamak Plasmas\/}.
\newblock Ph.D. thesis, University of York.

\bibitem{Hoelzl2014}
H{\"o}lzl M, Huysmans G, Merkel P, {\em et~al.\/} 2014 \enquote{Non-linear
  simulations of {MHD} instabilities in tokamaks including eddy current effects
  and perspectives for the extension to halo currents.}
\newblock {\em Journal of Physics: Conference Series\/} {\bf 561} 012011.

\bibitem{Fil2015}
Fil A, Nardon E, Hoelzl M, {\em et~al.\/} 2015 \enquote{Modeling a massive gas
  injection triggered disruption in {JET} with the {JOREK} code.}
\newblock {\em Physics of Plasmas\/} {\bf 22} 062509.

\bibitem{Liu2014}
Liu F, Huijsmans G, Loarte A, {\em et~al.\/} 2014 \enquote{Nonlinear {MHD}
  simulations of {QH}-mode plasmas in {DIII-D}.}
\newblock In {\em Proceedings of the 41st {EPS Conference on Plasma
  Physics}\/}. Berlin, Germany.
\newblock O5.135.

\bibitem{Strumberger2011}
Strumberger E, Merkel P, Tichmann C, and G{\"u}nter S 2011 \enquote{{Linear
  stability studies in the presence of 3D wall structures}.}
\newblock In {\em 38th EPS Conf. on Plasma Phys.\/}. ECA Vol. 35G P5.082,
  Strasbourg, France.

\bibitem{Wimp1984}
Wimp J 1984 \enquote{Computation with recurrence relations.}
\newblock {\em Pitman Publishing Inc., London\/}  305.

\bibitem{Wong1989}
Wong R 1989 \enquote{Asymptotic approximation of integrals.}
\newblock {\em Academic Press. Inc, San Diego\/}  432.

\end{thebibliography}
